\documentclass[a4paper,UKenglish,cleveref,autoref,thm-restate]{lipics-v2021}

\title{Reasoning about Quality in Hyperproperties} 

\author{Samuel Graepler}{Fakultät für Mathematik und Informatik, Universität Leipzig, Germany}{samuel.graepler@studserv.uni-leipzig.de}{}{}%

\author{Benjamin Monmege}{Aix Marseille Univ, CNRS,
   LIS, Marseille,
   France}{benjamin.monmege@univ-amu.fr}{https://orcid.org/0000-0002-4717-9955}{}%

 \author{Jean-Marc Talbot}{Univ. Bordeaux, CNRS, Bordeaux INP, LaBRI, UMR 5800, F-33400 Talence, France}{jean-marc.talbot@u-bordeaux.fr}{https://orcid.org/0009-0008-6536-5840}{}%

\authorrunning{S. Graepler, B. Monmege and J.-M. Talbot} 

\Copyright{Samuel Graepler, Benjamin Monmege and Jean-Marc Talbot} 

\ccsdesc[500]{Theory of computation~Logic and verification}

\keywords{Hyperlogics, Automata-based model checking, Quantitative verification}

\funding{This work was partially done while the first author was an intern at Aix-Marseille Université. This work was partly supported by ANR-23-PECL-0009 TRUSTINCloudS.}

\hideLIPIcs
\nolinenumbers

\usepackage{todonotes}
\usepackage{stmaryrd}
\usepackage{xspace, xparse}

\usepackage{soul}

\usepackage{tikz}
\usetikzlibrary{shapes.geometric,automata,positioning, arrows,quotes}
\tikzset{elliptic state/.style={ellipse,ellipse, minimum height=0.8cm,minimum
width=1.1cm, inner sep=0, outer sep=0,draw},fontscale/.style =
{font=\footnotesize}}

\renewcommand{\phi}{\varphi}
\newcommand{\N}{\mathbb N}
\newcommand{\Lang}{\mathcal{L}}
\newcommand{\K}{\mathcal{K}} 
\newcommand{\A}{\mathcal{A}}

\newcommand{\AP}{\ensuremath{\mathrm{AP}}\xspace}

\newcommand{\funs}{\mathcal{F}} 
\newcommand{\discfuns}{\mathcal{D}}

\newcommand{\LTL}{\ensuremath{\mathrm{LTL}}\xspace}
\newcommand{\LTLprop}{\ensuremath{\LTL_{\mathrm{prop}}}\xspace}
\newcommand{\LTLtemp}{\ensuremath{\LTL_{\mathrm{temp}}}\xspace}

\newcommand{\CTL}{\ensuremath{\mathrm{CTL}}\xspace}

\newcommand{\HyperLTL}{\ensuremath{\mathrm{HyperLTL}}\xspace}
\newcommand{\HyperLTLprop}{\ensuremath{\HyperLTL_{\mathrm{prop}}}\xspace}
\newcommand{\HyperLTLtemp}{\ensuremath{\HyperLTL_{\mathrm{temp}}}\xspace}
\newcommand{\HyperLTLtemppos}{\ensuremath{\HyperLTL_{\mathrm{temp}}^+}\xspace}
\newcommand{\HyperLTLtempneg}{\ensuremath{\HyperLTL_{\mathrm{temp}}^-}\xspace}
\newcommand{\HyperLTLtempforall}{\HyperLTLtemp^{\forall}}
\newcommand{\HyperLTLtempexists}{\HyperLTLtemp^{\exists}}

\newcommand{\Bool}{\mathrm{Bool}\xspace}

\newcommand{\inter}[1]{\llbracket #1 \rrbracket}

\newcommand{\sem}[2]{\llbracket #1 \rrbracket(#2)}
\newcommand{\hsem}[3]{\llbracket #1 \rrbracket_{#3}(#2)}

\newcommand{\chsem}[2]{\llbracket #1 \rrbracket_{#2}}

\DeclareMathOperator{\X}{\mathbf X} 
\newcommand\U{\mathbin{\mathbf U}}
\newcommand\R{\mathbin{\mathbf R}}
\DeclareMathOperator{\F}{\mathbf F} 
\DeclareMathOperator{\G}{\mathbf G}
\newcommand{\true}{\ensuremath{\mathrm{true}}\xspace}
\newcommand{\false}{\ensuremath{\mathrm{false}}\xspace}

\newcommand{\Traces}{\mathbb{T}}
\newcommand{\Lassos}{\mathbb{L}}

\newcommand{\TracesOf}[1]{\mathbb{T}(#1)}
\newcommand{\LassosOf}[1]{\mathbb{L}(#1)}

\newcommand{\scalmult}[1]{\ensuremath{\triangledown_{\!#1}}}
\newcommand{\lowequiv}[2]{\ensuremath{\mathrm{same}_\mathrm{low}(#1, #2)}}
\newcommand{\dummy}[1]{\ensuremath{\Lambda_\mathrm{high}(#1)}}
\newcommand{\ratio}[2]{\ensuremath{\mathrm{ratio}_\mathrm{low}(#1, #2)}}

\newcommand{\Discounts}{\ensuremath{D}}

\newcommand{\Valpositive}[3]{\ensuremath{V^{#1}_{#2, #3}}}

\begin{document}

\maketitle

\begin{abstract}
    Hyperproperties allow one to specify properties of systems that inherently involve not single executions of the system, but several of them at once: observational determinism and non-inference are two examples of such properties used to study the security of systems. Logics like HyperLTL have been studied in the past to model check hyperproperties of systems. However, most of the time, requiring strict security properties is actually ineffective as systems do not meet such requirements. To overcome this issue, we introduce qualitative reasoning in HyperLTL, inspired by a similar work on LTL by Almagor, Boker and Kupferman \cite{Almagor14} where a formula has a value in the interval $[0, 1]$, obtained by considering either a propositional quality (how much the specification is satisfied), or a temporal quality (when the specification is satisfied).
    We show decidability of the approximated model checking problem, as well as the model checking of large fragments.
\end{abstract}

\section{Introduction}

Software safety and security have become a critical concern. Hence, ensuring that applications are free from safety issues or security vulnerabilities is an important area of the verification process. Model checking \cite{HMC2020} is a successful formal method to automatically verify whether a system meets a specification. This specification is often described as formulas from dedicated logics, for instance \LTL to express properties on execution traces of a system \cite{SisCla85} regarding safety or \CTL to express properties on the tree structure of possible executions of the system~\cite{ClarkeEmerson81}.  Model checking, in particular of \LTL, has been widely studied and leads to several tools like SPIN \cite{Spin} and Spot \cite{Spot}. 

Over the last years, hyperlogics have been found out to be another convenient tool to express properties about systems. These logics define hyperproperties \cite{Hyperproperties} that consider not single executions as \LTL, but several of them at once. Hence, hyperlogics are often used to express properties relating multiple
executions of a system, e.g.,
network properties like congestion \cite{7588075}, or security and information-flow properties \cite{info-flow,10.1007/11547662_24,601322}. Let us focus here on the security property of non-interference \cite{GogMes82}, which holds when executions of systems do not reveal sensitive information about manipulated data to malicious observers: secret information cannot be deduced from information of low sensitivity exposed by the system executions. As running examples, we consider two desirable properties to meet such requirements. We express them in the logic \HyperLTL which extends \LTL with a prefix of quantifications over traces variables~\cite{ClarkFink14}. First, the property called \emph{observational determinism}~\cite{DBLP:conf/csfw/ZdancewicM03} expresses the fact that when any two executions start with the same information of low sensitivity, then when observed all along the computation, the two traces cannot be distinguished on these low information (the system behaves deterministically with respect to them). To describe this property, we write a  \HyperLTL formula $\varphi_{\mathsf{OD}}$ 
\begin{equation}
 \varphi_{\mathsf{OD}} \quad = \quad   \forall\pi\, \forall\pi' \quad \lowequiv \pi {\pi'} \ \rightarrow \ \G \lowequiv \pi {\pi'} 
\end{equation}
which  uses the \LTL formula $\lowequiv \pi {\pi'}$, with two free trace variables, that states that at the first instant, $\pi$ and $\pi'$ fulfil the same set of low atomic propositions. 
The ``Globally'' operator $\G$ guarantees that this atomic property $\lowequiv \pi {\pi'}$ holds for any synchronized suffixes of this pair of traces, and thus, at any time.

The second property called \emph{non-inference} \cite{DBLP:conf/sp/McLean94} expresses the fact that high sensitive information are not exposed through low level one: for any trace, there must exist another one with arbitrary or dummy high sensitive information but with similar low-level information all along the computation. This property can be written as a  \HyperLTL formula as:
\begin{equation}
\varphi_{\mathsf{NI}} \quad = \quad \forall\pi\, \exists\pi'\quad \G\dummy{\pi'}\land \G\lowequiv \pi {\pi'}  
\end{equation}
where the formula $\dummy {\pi'}$ is an \LTL formula checking that the only high (i.e.~non-low) atomic propositions are a dummy one.

The model checking problem for  \HyperLTL which aims to show that the property holds for the set of all execution traces of a system is decidable 
\cite{ClarkFink14,DBLP:conf/vmcai/Finkbeiner21}.    

Unfortunately, most of the time, requiring strict security properties is
actually ineffective as programs do not meet such drastic requirements. To overcome
this issue, it has been proposed to quantify security breaches as, e.g., in
quantitative information flow \cite{minentropy}. Such approaches allow some
information leaks provided they are acceptable, i.e.~of amplitude not beyond
some threshold; they have been studied in the framework of hyperproperties and
hyperlogics~\cite{DBLP:journals/tcs/YasuokaT14,DBLP:journals/corr/abs-1905-13514}, in particular by considering counting extensions of \HyperLTL 
where the number of different models of a formula can be compared with some threshold.

In this article, we follow a different approach 
and base our work on the extensions of \LTL proposed in \cite{Almagor14}. 
In this latter, the authors proposed a framework enriching the syntax and semantics of LTL for reasoning about the \emph{quality}  of systems;  the semantics of a formula is a value in $[0,1]$ reflecting the quality of a trace, that is, at which extent it is satisfied by this trace (0 when the formula is not satisfied at all and 1 when the formula is fully satisfied). 
Two ways on how to reason about quality in the context of LTL are explored there. The first one, the ``propositional
quality'',  focuses on the different ways \emph{how} a specification can be satisfied. It is embodied by \LTLprop, an extension of $\LTL$ with logical operators defined as mappings over $[0,1]$. The second one, the ``temporal quality'', focuses on \emph{when} eventualities are satisfied and the corresponding $\LTL$ extension is 
\LTLtemp. This latter considers temporal operators with discount which makes it possible to assign different values to a formula depending on \emph{when}
certain eventualities happen\footnote{We
use different names for the two logics, to be consistent with the hyperlogics
we introduce afterwards.}.

We lift these qualitative extensions of \LTL to the setting of hyperproperties. In the Boolean setting, the models of a hyperproperty (e.g., a \HyperLTL formula) are sets of traces. Hence, a \HyperLTL formula associates a Boolean value with each set of traces.
To incorporate quality, a formula thus must map each set of traces to a value in $[0, 1]$. We propose qualitative extensions of \HyperLTL, similar to the two logics \LTLprop and \LTLtemp with respect to $\LTL$, named respectively \HyperLTLprop and \HyperLTLtemp. For the former, coming back to our example above, the condition on observational determinism can be relaxed  by requiring the difference on low information to remain ``small'' all along the computation: a formula of \HyperLTLprop thus allows one to associate with a given system a propositional quality describing guarantees about observational determinism. One may also consider as acceptable two traces that finally differ quite far from their origin: this is enabled in \HyperLTLtemp.

A similar attempt has been achieved in \cite{FerLuo24} to lift \LTLprop to the setting of \emph{alternating-time temporal logic} to be able to quantify over strategies of agents in a multi-agent system: the same kind of fuzzy functions are added to talk about the quality in strategic reasoning.

Our main objective is to study the model checking of the hyperlogics \HyperLTLprop and \HyperLTLtemp. It consists in deciding whether the semantics of a (closed) formula is above (or below) a given threshold $v\in [0,1]$ when the set of traces in consideration is the one generated by a system (a Kripke structure). Qualitative model checking will therefore indicate to which extent the set of traces of a Kripke structure satisfies a formula.

\HyperLTLprop is studied in Section~\ref{sec:propHyperLTL}. We show that the semantics of a \HyperLTLprop formula can take only a finite set of values in $[0, 1]$. 
We use this crucial property to design an automata-based model checking algorithm by building a non-deterministic Büchi automaton (NBA) from a \HyperLTLprop formula for which we test emptiness. This finiteness property also implies that  \HyperLTLprop is not more expressive than  \HyperLTL. 

For \HyperLTLtemp, studied in Section~\ref{sec:hyperLTL_disc[D]}, the situation is much more delicate as the set of possible values of a formula is no longer finite. 
We develop nonetheless automata-based techniques for model checking following two directions. We first solve the model checking problem with promise for \HyperLTLtemp that is correct 
when the semantic value of the \HyperLTLtemp formula is different from the threshold $v$. 
We then consider fragments of \HyperLTLtemp for which we solve the full model checking problem: the first fragment allows arbitrary quantifier prefixes but limits the interplay between negations and discounted operators. We show that in this case, the set of possible values, although infinite, enjoys a nice structure. This allows us to reduce this problem to the one with promise but in which the promise turns out to be always satisfied. For the second fragment, one considers arbitrary quantifier-free parts of formulas but forbid quantifier alternation. 
For these two fragments, we provide examples showing that they are still sufficient to express interesting hyperproperties on the quality of systems. 

We describe our results in a weighted setting where systems (Kripke structures) that generate
traces associate with each time step a mapping giving to atomic propositions a weight in $[0,1]$, instead of just the subset of atomic propositions that currently hold.
This allows one to incorporate
more meaningful qualitative reasoning since basic weights come from the system
description itself. For instance, weights can encode costs, time delays or probabilities. 
A similar setting for \LTL has been considered in~\cite{FaeLeg08}. 

Other kinds of logic than temporal logic, like HyperFO \cite{DBLP:conf/stacs/Finkbeiner017} that is a hyper-extension of the first-order logic, could also be extended with quality. However, as advocated in \cite{Almagor14}, weighted extensions of logics like FO (or even MSO, i.e.~monadic second order logic) such as wMSO \cite{DBLP:journals/tcs/DrosteG07} or wFO \cite{DBLP:journals/tocl/BolligGMZ14} are not very suitable to define specifications due to their undecidable nature. Hence, we believe that a weighted extension of HyperFO \cite{DBLP:conf/stacs/Finkbeiner017} would not fit our goals.

The present article is a long version of the article published at the conference CSL~2026~\cite{GraMon26}. This latter has been extended with: 
\begin{itemize}
    \item full and updated proofs, 
    \item some new examples, 
    \item a rewriting of what we called \emph{approximate model checking} of \HyperLTLtemp, into what we now call a \emph{promise model checking} that does not require incorporating any approximate factor,
    \item the proof that for any positive $\Delta$, the semantic of any \HyperLTLtemp formula can be approximated by the semantic of some \HyperLTLprop formula modulo a $\Delta$-term. 
     This yields as a by-product an approximate model checking algorithm for \HyperLTLtemp.
\end{itemize}

\section{Preliminaries}
\label{sec:preliminares}

\newcommand\weights{\ensuremath{\mathbb W}\xspace}

\subparagraph*{Kripke structures}
We let $\AP$ be a finite set of atomic propositions. We
consider \emph{traces} that are infinite sequences of atomic propositions each
weighted by a weight in $[0,1]$, i.e.~traces are elements of $([0,1]^{\AP})^\omega$. We
denote $\Traces$ the set of all traces, and call $\Traces_\weights$ the subset
of traces where all weights are taken from a finite set $\weights\subset [0,1]$.
A particular case is when $\weights=\{0,1\}$, in which case we recover the
classical traces, and we call such traces \emph{Boolean traces} in the
following. Amongst all traces, \emph{lassos} are the ones of the form $u
v^\omega$ with $u,v \in  ([0,1]^{\AP})^*$ (a finite word, followed by the
infinite repetition of some finite word). We denote $\Lassos$ the set of all
lassos, and $\Lassos_\weights$ if weights are restricted in the finite set $\weights$. 
For a trace $t\in \Traces$, we write~$t[i]$ to refer to the $i$-th element in
$t$, for $i\in\N$. We denote by $t[i,\infty]$ the suffix trace starting at the $i$-th element. 

A \emph{(weighted) Kripke structure} is a tuple $\K=\langle
S, \weights, I,\rightarrow,L\rangle$, where $S$ is a finite set of states,
$\weights\subset [0,1]$ is a finite set of weights, $I\subseteq S$ is the set
of initial states, $\rightarrow \ \subseteq S\times S$ is a total transition
relation, and $L\colon S\rightarrow \weights^{\AP}$ is a labelling function,
assigning a weight to each atomic proposition in each state. The weight can
encode for example costs, time delays or probabilities for the atomic
propositions to be true. In case the weights are only $0$ and $1$, we recover
the classical Kripke structures used, e.g., in the model checking of \LTL: in this article, we call such structures \emph{Boolean Kripke structures}.

A \emph{path} $s$ in $\K$ is a sequence $s=s_0 s_1 \cdots$
with $s_0 \in S$ and $s_i\rightarrow s_{i+1}$ for all $i\in \mathbb{N}$. Since
$\rightarrow$ is total, there are no deadlocks and thus all paths are of infinite length. The set of all
\emph{traces} in $\K$, denoted $\TracesOf \K$, is defined as the set of traces
$t\in \Traces_\weights$ of the form $t=L(s_0) L(s_1) \cdots$ for a path $s_0
s_1\cdots$ in $\K$. $\LassosOf \K$ denotes the set of all lassos of $\K$, that
is, traces from $\TracesOf \K$ that are lassos.

\subparagraph*{Büchi automata}
The model checking of \LTL as well as \HyperLTL, consisting in checking whether a given (Boolean) 
Kripke structure satisfies a given closed formula, relies on automata-based techniques. We recall here the automata used in this context. 

A \emph{non-deterministic Büchi automaton} (NBA) is a tuple
$\A=\langle\Sigma,Q,Q_0,\delta,F\rangle$, where $\Sigma$ is a finite
alphabet, $Q$ is a finite set of \emph{states}, $Q_0\subseteq Q$ is the set of
\emph{initial states}, $\delta:Q\times\Sigma\rightarrow 2^Q$ is a
\emph{transition function} and $F\subseteq Q$ is the set of 
\emph{accepting states}. An infinite sequence $r=r_0 r_1 \cdots$  of states is called a
\emph{run} on a word $w=w_1 w_2 \cdots \in\Sigma^\omega$ if $r_0\in Q_0$ and for
every $i\geq0$, $r_{i+1}\in\delta(r_i,w_{i+1})$. An NBA
\emph{accepts} a word if there exists a run $r$ on it that visits $F$  infinitely often, 
i.e.~for $\inf(r)=\{q \mid r_i=q \text{ for infinitely many } i\in\N\}$, it holds that $\inf(r)\cap F\neq \emptyset$. 
The accepted language of an NBA, denoted by $\Lang(\A)$,
is the set of words accepted by $\A$. 

The emptiness problem of an NBA, 
i.e.~deciding whether the accepted language is empty, is decidable in linear
time and is $\mathrm{NLOGSPACE}$-complete \cite{ltl-modelchecking}. 

Given a Kripke structure $\K=\langle S, \weights, I,\rightarrow,L\rangle$
defined over a set $\AP$ of atomic propositions and a finite set of weights $\weights$, one can define an NBA 
$\A_\K$ over the alphabet $\weights^{\AP}$ accepting the language~$\TracesOf \K$ as 
$(\weights^{\AP}, S \cup \{\mathsf{init}\}, \{\mathsf{init}\} , \delta_\K, S)$ where 
for all $q \in S \cup \{\mathsf{init}\}$ and all $m \in \weights^{\AP}$,
$$\delta_\K(q,m)  = \begin{cases}
\{q' \mid q' \in I, L(q') = m \} & \text{if $q= \mathsf{init}$}\\
\{ q' \mid q' \in S, q \rightarrow q', L(q') = m  \}  & \text{otherwise}
\end{cases}
$$

\subparagraph*{HyperLTL}
The logic  \HyperLTL, a hyperlogic extension of Linear Time Logic (\LTL) was  introduced in~\cite{ClarkFink14}. It extends \LTL with universal and existential quantifiers over traces allowing to express properties on several traces. 
Models for \HyperLTL formulas are then sets of traces and not individual traces. Hence, the set of all models of a \HyperLTL formula is a set of sets of traces, a hyperproperty.

\begin{definition}
\label{def:hyperLTL}
Let $\mathcal{V}$ be the infinite set of (trace) variables. The \emph{formulas of \HyperLTL} are defined by the following grammar,   for all atomic propositions $p\in \AP$ and variables $\pi\in\mathcal{V}$:
    \begin{align*}
            \psi& ::= \exists\pi\, \psi \mid \forall\pi\, \psi \mid \phi \\
            \phi& ::= p_\pi \mid \neg\phi \mid \phi \vee \phi \mid \X\phi \mid \phi\U\phi
    \end{align*}
\end{definition}    
    
   As usual, additional operators can be derived : $\phi_1\wedge\phi_2 =\neg((\neg\phi_1)\vee(\neg\phi_2))$, $\true =p_\pi \vee \neg p_\pi$ and $\false =\neg \true$, $\phi_1 \rightarrow\phi_2 =\neg\phi_1\vee\phi_2$, $\phi_1\leftrightarrow\phi_2 =(\phi_1 \rightarrow\phi_2)\vee(\phi_2 \rightarrow\phi_1)$, the \emph{Finally} operator $\F\phi =\true\U\phi$, and the \emph{Globally} operator $\G\phi =\neg\F(\neg\phi)$. 
    
If $\psi=Q_1\pi_1\cdots Q_n\pi_n\phi$ for $Q_i\in\{\exists,\forall\}$ for $i\in\{1,\dots,n\}$ and $\phi$ quantifier-free, we call $\phi$ the
\emph{quantifier-free part} of $\psi$ and the \emph{quantifier depth} $\psi$ is $n$. 

The \emph{alternation depth} $\alpha(\psi)$ of a $\HyperLTL$ formula $\psi$ is defined as the maximal number of quantifier alternations from existential to universal, or vice versa, in a branch of the  tree structure of $\psi$. Formally, $\alpha(\psi) = \max(\alpha_\exists (\psi),  \alpha_\forall (\psi))$ where two new measures $\alpha_\exists$ and $\alpha_\forall$ are co-recursively defined on the structure of the formula by: 
    $$\begin{array}{l}
        \alpha_\exists(\psi_0) =  \alpha_\forall(\psi_0)  = 0 \ \text{ if } \psi_0 \text{ is quantifier-free }\\
        \alpha_\exists(\exists \pi\, \psi_0) =  \alpha_\exists(\psi_0) \qquad \quad    \alpha_\forall(\forall \pi\, \psi_0)  = \alpha_\forall(\psi_0) \\
        \alpha_\exists(\forall\pi\, \psi)  = \alpha_{\forall}(\psi_0)+1  \qquad \alpha_\forall(\exists \pi\, \psi)  = \alpha_{\exists}(\psi_0)+1
      \end{array}$$  

The semantics of \HyperLTL formulas over (weighted) traces slightly 
differs from the classical semantics of \HyperLTL which associates with each
formula a set of traces. However, the two definitions coincide in the case of
Boolean Kripke structures. 

\begin{definition}
\label{def:HyperLTLsem}
Given a trace assignment $\Pi$ over $\Traces$, mapping trace variables from $\mathcal{V}$ to traces
from $\Traces$, and a non-empty set of traces $T \subseteq \Traces$
used as a ranging set for quantified variables\footnote{Notice that we do not
require $\Pi$ to map variables to the restricted set of traces $T$.}, the
semantics $\hsem\psi \Pi T$ of a formula~$\psi$, called the
\emph{satisfaction value} of $\psi$ over $T$ given~$\Pi$, is the weight in $[0,1]$ defined
recursively as:
\begin{align*}
    \hsem{p_\pi} \Pi T & = \Pi(\pi)[0](p) \\
    \hsem {\lnot\phi}\Pi T  &= 1-\hsem \phi \Pi T \\
            \hsem {\phi_1 \lor \phi_2}\Pi T  & =\max(\hsem{\phi_1}\Pi T,\hsem {\phi_2}\Pi T) \\
    \hsem{\X\phi} \Pi T  &= \hsem\phi {\Pi[1,\infty]} T\\
    \hsem{\phi_1 \U \phi_2}\Pi T & = \makebox[0pt][l]{$\displaystyle\sup_{i\geq 0}\Big(\min\big(\hsem{\phi_2} {\Pi[i,\infty]} T,\min_{0\leq j < i}\hsem{\phi_1}{\Pi[j,\infty]} T\big)\Big)$} \\     \hsem{\forall\pi\, \psi} \Pi T & = \inf_{t\in T}\hsem\psi{\Pi[\pi\mapsto t]} T   \\
    \hsem{\exists\pi\, \psi}\Pi T &= \sup_{t\in T}\hsem\psi{\Pi[\pi\mapsto t]} T
\end{align*}
where $\Pi[\pi\mapsto t]$ is the trace assignment identical to $\Pi$ except that it maps $\pi$ to $t$, 
and $\Pi[i,\infty]$ denotes the suffix trace assignment, defined for all variables $\pi$ as $\Pi[i,\infty](\pi)=\Pi(\pi)[i,\infty]$. 
\end{definition}

Note that in case of a trace assignment that only contains Boolean traces (that we call \emph{Boolean assignments}), it is easy to see that the semantics of \HyperLTL formulas coincides with the classical semantics of \HyperLTL, i.e.~maps all Boolean assignments to $0$ or $1$ such that the set of assignments mapped to $1$ are the ones satisfying the formula.

As one may expect, for the existentially quantified formula $\exists\pi\, \psi$,  when $\psi$ in which $\pi$ is set to some trace $t$ is evaluated to $1$, then so is $\exists\pi\, \psi$. Hence, existential quantification is intuitively a disjunction over the set of all traces from T and thus, corresponds to a supremum. Universal quantification is defined dually as an infimum.  

The semantics of the derived operators 
$\F$ and $\G$ are, therefore, given by   
$$\hsem{\F\phi}\Pi T=\sup_{i\geq0}\hsem \phi{\Pi[i,\infty]} T \quad  \text{and} \quad  
\hsem{\G\phi}\Pi T=\inf_{i\geq0}\hsem \phi{\Pi[i,\infty]} T$$.
For a closed formula $\psi$ (i.e.~without free variables), the satisfaction value of $\psi$ over a Kripke structure $\K$ is the value $\hsem \psi{[ \, ]} {\TracesOf \K}$ 
(denoted simply $\chsem \psi {\TracesOf \K}$ from now on) taken over the empty assignment $[ \, ]$.

\subparagraph*{HyperLTL, LTL and synchronous product of Kripke structures}

Let us consider, for $n \in \mathbb{N}$, an ordered sequence $ \pi_1,\pi_2, \ldots, \pi_n $ of variables from $\mathcal{V}$ and $\AP$ a finite set of atomic variables.

Let us denote by $\AP_i = \{p_{\pi_i} \mid p \in  \AP \}$ the set of atomic propositions indexed by the trace variable $\pi_i$. We then let $\AP_{\le n}$ be the set $\bigcup_{1 \le i \le n} \AP_{i}$. Notice that $\AP_{\le 0}$ is empty. 

\begin{definition} 
\label{def:tPI}
Let $\Pi = [\pi_1\mapsto t_1,\dots,\pi_n\mapsto t_n]$ be a trace assignment with $t_i \in \Traces$ for all $1 \le i \le n$ 
The trace $t_\Pi$ induced by $\Pi$ is an element of $(\weights^{\AP_{\le n}})^\omega$ and is defined for all $k \in \mathbb{N}$, for all $1 \le i \le n$, for all $p \in \AP$, by
$(t_\Pi[k])(p_{\pi_i}) = (t_i[k])(p)$. 
\end{definition}


For a Kripke structure $\K=\langle S, \weights, I,\rightarrow,L\rangle$, we define $\K^i$ the synchronous product of $i$ copies of  
$\K$. 

\begin{definition}
\label{def:productstruct} For all $i \in \mathbb{N}$,  $\K^i$ is a Kripke structure over the set $\AP_{\leq i}$ of atomic propositions, defined inductively as follows: 
\begin{itemize}
\item $\K^0 = \langle \{s\}, \{0\}, \{s\}, \{(s,s)\}, L\rangle$ for some arbitrary state $s$, with the labelling $L(s)$ being empty (since there are no atomic propositions in $\AP^{\le 0}$);
\item assuming $\K^i = \langle S_i, \weights, I_i,\rightarrow_i,L_i\rangle$, $\K^{i+1}$ is the Kripke structure 
\[\langle S_i \times S, \weights, I_i \times I, \{((s_{i},s),(s'_{i},s')) \mid s_{i}\rightarrow_i s'_{i}, \  s \rightarrow  s'\}, L_{i+1}
\rangle\]
where $L_{i+1}((s_i,s))(p_{\pi_j}) =  L_{i}(s_i)(p_{\pi_j})$ if $j\le i$ and $L_{i+1}((s_i,s))(p_{\pi_{i+1}}) = L(s)(p)$. 
\end{itemize}
\end{definition} 

It is easy to see that 
\begin{lemma}
\label{lem:traceproduct}
Let $\K$ be a Kripke structure and $\Pi = [\pi_1\mapsto t_1,\dots,\pi_n\mapsto t_n]$ be a trace assignment. Then $t_i \in \TracesOf \K$ for all $1 \le i \le n$ iff  
the trace $t_\Pi$ induced by $\Pi$ is an element of  $\TracesOf {\K^n}$.
\end{lemma}

\begin{remark}
\label{rem:HyperLTLtoLTL}
Let $\psi = Q_1 \pi_1 \ldots Q_n \pi_n\, \phi$ be a closed \HyperLTL formula, $\phi$ being its quantifier-free part with $n$ free variables and 
$Q_i\in\{\exists,\forall\}$ for all $i\in\{1,\dots,n\}$. The formula $\phi$ turns out to be a formula of the logic \LTL defined over the set of atomic propositions $\AP_{\le n}$.

Following \cite{FaeLeg08}, the semantics of LTL formulas is given by a mapping from traces of $(\weights^{\AP_{\le n}})^\omega$ to $[0,1]$ 
such that $\sem \phi {t_\Pi} = \hsem \phi \Pi T$, for arbitrary $T$ (as the  \HyperLTL formula $\phi$ is non-quantified). 
\end{remark}

\section{Propositional quality for \HyperLTL}
\label{sec:propHyperLTL}

\subsection{Definitions and  examples}  \label{sec:prop-Defs}

In the same way that \HyperLTL is built upon \LTL, \HyperLTLprop corresponds to the hyperlogic extension of the linear temporal logic \LTLprop, for \emph{LTL with propositional quality}~\cite{Almagor14}.\footnote{This logic is called $\LTL[\funs]$ in~\cite{Almagor14}.}
This logic generalises LTL by replacing the Boolean operators by a set $\funs$ of arbitrary functions over $[0,1]$. Hence, the syntax of 
\LTL is enriched by the construct $f(\phi_1,\ldots,\phi_k)$ for $f\in\funs$,  the temporal operators remaining the same. 
Strictly speaking, it is then a family of logics, parameterized by $\funs$. 
Hence, beyond classical Boolean operators ($\lor, \land, \lnot$) that can obviously be defined as such functions (and that we thus suppose to always be members of $\funs$), the set $\funs$ can contain quantitative functions like $\oplus_\alpha$ and $\scalmult\alpha$ for $\alpha\in [0,1]$, that associates with $x$ and $y$ the weighted sum $x\oplus_\alpha y=\alpha x+(1-\alpha)y$ and the scalar multiplication $\scalmult\alpha x=\alpha x$. 
Intuitively, the functions in $\funs$ allow one to give weights to subformulas,  i.e.~parts of our specification, to moderate their ``effect'' in the formula where they occur. Let us point out that with such operators the semantics for formulas indeed belongs to the interval $[0,1]$.


\begin{definition} \label{def:HyperLTL[F]} 
Assuming $\funs$ is a finite set of function symbols,  the syntax of \HyperLTLprop is given by 
\begin{align*}
    \psi & ::= \forall\pi\, \psi \mid \exists\pi\, \psi \mid \phi  \\
    \phi & ::= \true \mid \false \mid p_\pi \mid f(\phi,\dots,\phi)\mid \X\phi \mid \phi \U \phi
\end{align*}
for all $p\in \AP$, $\pi\in\mathcal V$, and $f\in\funs$.
These functions $f$ being interpreted as mappings from $[0,1]^k$ to $[0,1]$ (for functions of arity $k$), the \HyperLTL semantics is extended for 
\HyperLTLprop as 
\[\hsem{f(\phi_1,\dots,\phi_k)} \Pi T  = f(\hsem{\phi_1}\Pi T,\dots,\hsem{\phi_k}\Pi T)\]
for all trace assignments $\Pi$ over $\Traces$, and non-empty sets of traces $T \subseteq \Traces$. 
\end{definition}

We give two examples showing that this extension of \HyperLTL with propositional quality allows to model hyperproperties with a very different flavour than other quantitative extensions of \HyperLTL (like counting extensions considered in \cite{DBLP:journals/corr/abs-1905-13514}). The two first examples deal with Boolean Kripke structures, while the last one is a more general example that can be applied to a weighted Kripke structure.

\begin{example}\label{exmpl:obs det hyperltl[F]} 
We propose some quantitative
versions of the \emph{observational determinism} property presented in the introduction. First, let us formally define the subformula 
$\lowequiv \pi {\pi'}$ of the formula  $\varphi_{\mathsf{OD}}$, having two free variables $\pi$ and $\pi'$, as 
$\bigwedge_{p\in \mathrm{low}} p_\pi \leftrightarrow p_{\pi'}$, where $\mathrm{low}$ is the subset of "low sensitivity" atomic propositions. To compute the propositional quality of observational determinism, we adapt this formula by using the $\oplus$ weighted function 
in order to express the ratio of the agreement of two traces at the current step: 
$$\ratio \pi {\pi'} := \bigoplus_{p\in \mathrm{low}}\frac{1}{|\mathrm{low}|}(p_\pi \leftrightarrow p_{\pi'})$$
Then, the formula $\G \ratio \pi {\pi'}$ aggregates these values at every step by considering an infimum. Thus, we can modify formula 
$\varphi_{\mathsf{OD}}$ to compute the smallest total ratio separating two traces of the system that agree on the first step, thus relaxing the Boolean observational determinism:  
\[\forall\pi\, \forall\pi'\quad \lowequiv{\pi}{\pi'} \to \G \ratio \pi{\pi'}\] 
One may use the interval $[0,1]$ in a less strict way and split it into several pieces, each of them encoding different ``truth'' status of the specification. For instance, we can
refine this formula by separating the values $1$ obtained when the system non trivially fulfils observational determinism from the degenerated case where there are no two different traces which agree on the low atomic propositions on the first step:
 \[\varphi_{\mathsf{OD}}^{\mathrm{prop}} \quad = \quad 
 \forall\pi\, \forall\pi'\quad \scalmult{\frac{1}{2}}\G 
 \ratio \pi {\pi'}
  \lor \scalmult{\frac{3}{4}}(\lnot\lowequiv \pi {\pi'}\lor \pi= \pi')\]
  where the subformula $\pi = \pi'$ is used as a shortcut for the Boolean formula $\G(\bigwedge_{p\in \AP} p_\pi \leftrightarrow p_{\pi'})$, that states on Boolean Kripke structures 
  that for every atom $p$ and all positions,  the truth value of $p$ is the same in $\pi$ and $\pi'$ at this position.
  Here we get half of the \emph{smallest ratio of corresponding low atomic propositions} (and thus a value in the interval $[0, 1/2]$), if there are two different traces which agree on the low atomic propositions of the
  first step, or $\frac{3}{4}$ (any value outside $[0,1/2]$ could be convenient) if there are no such traces. The universal quantifications in the formula correspond to some minimisation and thus enable to consider the worst-case scenario.
We could also consider the formula 
\[\forall\pi\, \forall\pi'\quad \lowequiv{\pi}{\pi'} \to \F \G (\ratio \pi{\pi'})\]
with the addition of an operator $\F$ to compute a $\lim\sup$ ratio, allowing for traces to vary a lot at the beginning, as long as they become close on the long term.
\qed
\end{example}

\begin{example}
    We present also
a modification of the \emph{non-inference} property $\varphi_{\mathsf{NI}}$ from the introduction. First, the subformula $\dummy{\pi'}$ can be described as $\lambda_{\pi'}\land \bigwedge_{p\notin\mathrm{low}\cup\{\lambda\}} \lnot p_{\pi'}$. Then, to describe the propositional quality of non-inference, we could ask whether, for every trace, we can find another trace with $\lambda$ as the only high atomic proposition, but which behaves \emph{almost} the same for a low user. This approximation in the quality can be obtained by using a new weighted function $\mathrm{threshold}_{>k} (x)$, with $k\in [0,1]$ that equals $1$ if $x$ is greater than $k$, and $0$ otherwise. The non-inference can then be relaxed as follows: 
\[\varphi_{\mathsf{NI}}^{\mathrm{prop}} \quad = \quad 
\forall\pi\, \exists\pi'\quad \G \dummy{\pi'}\land \mathrm{threshold}_{>k}(\ratio \pi{\pi'})\]
The semantics of the formula is Boolean, though it uses weighted functions to define it.   \qed
\end{example}

\begin{example}
    As a last example, we demonstrate what happens when we deal with a weighted Kripke structure that maps every atomic proposition to a probability that it holds in a given state. We can replace the formula $\ratio \pi {\pi'}$ of Example~\ref{exmpl:obs det hyperltl[F]} by its probabilistic version: 
    $\bigwedge_{p\in \mathrm{low}} f(p_\pi, p_{\pi'})$, where $f$ is the mapping defined for $x, y\in [0,1]$ by $f(x,y) = xy+(1-x)(1-y)$, giving the probability that $p_\pi$ and $p_{\pi'}$ are the same, $x$ being here the probability of $p_\pi$, and $y$ of $p_{\pi'}$. By considering the formula $\G \big(\bigwedge_{p\in \mathrm{low}} f(p_\pi, p_{\pi'})\big)$, we thus compute the smallest probability over all steps that low atomic propositions coincide. 
    The formula below thus computes the smallest such probability over any two traces: 
\[\forall\pi\, \forall\pi'\quad \lowequiv{\pi}{\pi'} \to \G \Big(\bigwedge_{p\in \mathrm{low}} f(p_\pi, p_{\pi'})\Big)\] 
    If the semantics of this formula is $1$, the system fulfils observational determinism. If the semantics of this formula is $0$, the system contains two traces where a certain proposition $p$ differs with probability $1$ at some position. The values between $0$ and $1$ give a probabilistic quality of observational determinism: value $1/2$, e.g., means that in any two traces, at any position, and for any low atomic proposition, the probability that the proposition coincides is at least~$1/2$.\qed
\end{example}

\subsection{A finite-valued property}

We fix in this section a finite set $\weights$ of weights in $[0,1]$.
Even though the family $\funs$ may contain functions ranging over the whole interval $[0,1]$, it turns out that every \HyperLTLprop formula~$\psi$
enjoys a finite-valued property: considering $V_\weights(\inter{\psi})$, the set of possible satisfaction values of the formula $\psi$, that is 
$$V_\weights(\inter{\psi}) = \{\hsem \psi \Pi T \mid \Pi \colon \mathcal{V} \rightharpoonup \Traces_\weights,\; T \subseteq \Traces_\weights \text{ non-empty}\}$$
it can be shown that this set is finite as one can compute a finite over-approximation of it. 
This has been shown for \LTLprop in~\cite[Lemma~2.5]{Almagor14} (when $\weights=\{0,1\}$), and we extend the proof here to \HyperLTLprop and arbitrary finite set $\weights$; this property will be used afterwards in our model checking algorithm.

\begin{definition} For any \HyperLTLprop formula  $\psi$ defined over $\funs$, the set $V_\weights(\psi)$ included in  $[0,1]$ is defined inductively by:
\begin{align*}
    &V_\weights(p_\pi)=\weights\cup\{0,1\} \qquad V_\weights( \X \phi) = V_\weights(\phi) \qquad V_\weights(\phi_1 \U \phi_2) = V_\weights(\phi_1) \cup V_\weights(\phi_2) \\ 
    &V_\weights(f(\phi_1,\dots,\phi_k)) =  \{f(v_1,\dots,v_k) \mid v_1 \in V_\weights(\phi_1), \ldots,  v_k \in V_\weights(\phi_k), f \in \funs \}\\
    &V_\weights(\forall\pi\, \psi) = V_\weights(\exists\pi\,\psi)=V_\weights(\psi)
\end{align*}  
for $p\in \AP$, $\pi\in \mathcal V$, $f \in \funs$.
\end{definition}

This set $V_\weights(\psi)$ turns out to be finite as, essentially, the number
of approximated satisfaction values only depends on $\#\psi$, the number of
occurrences of atomic propositions in the formula $\psi$. For instance, $\#
(p_\pi \U f(p_\pi, q_\pi, p_{\pi'})) = 4$ since the formula uses atomic
propositions $p_\pi$ (twice), $q_\pi$ and $p_{\pi'}$. Formally, 
 
\begin{restatable}{lemma}{lemnumbersatisfacvalues}
\label{lem:number_satisfac_values}
For all formulas $\psi$ of \HyperLTLprop, $V_\weights(\inter{\psi}) \subseteq V_\weights(\psi)$ and $|V_\weights(\psi)|\leq |\weights|^{\#\psi}$.
\end{restatable}

\begin{proof}
    The proof goes by induction on the formula $\psi$.
\begin{itemize}
    \item For an atomic proposition, $V_\weights(\inter{p_\pi}) =\weights= V_\weights(p_\pi)$ and $\#p_\pi=1$ and $|V_\weights(p_\pi)|=|\weights|\leq |\weights|^1$.
    \item If $\psi=f(\phi_1,\dots,\phi_k)$ with $f\in \funs$,
    by definition, $V_\weights(\inter{\psi}) = f(V_\weights(\inter{\phi_1}),\dots,V_\weights(\inter{\phi_k}))$. As by induction hypothesis, for all $1 \le i \le k$, we have $V_\weights(\inter{\phi_i}) 
    \subseteq V_\weights(\phi_i)$,  $V_\weights(\inter{\psi}) \subseteq V_\weights(f(\phi_1,\dots,\phi_k))$. To bound $|V_\weights(\psi)|$, notice that the
    number of possible inputs for $f$ in the definition of $V_\weights(\psi)$ is at most $\prod_{i=1}^k |V_\weights(\phi_i)|$,
    thus this is also an upper bound for the outputs. By induction hypothesis
    $|V_\weights(\phi_i)|\leq |\weights|^{\#\phi_i}$ for all $i \in \{1,\dots,k\}$, so we obtain
    $|V_\weights(\psi)|\leq \prod_{i=1}^k |V_\weights(\phi_i)| \leq \prod_{i=1}^k |\weights|^{\#\phi_i} =
    |\weights|^{\sum_{i=1}^k \#\phi_i} = |\weights|^{\#\psi}$.
    \item If $\psi = \X\phi$ then by definition, $V_\weights(\inter{\psi}) \subseteq V_\weights(\inter{\phi})$ and $V_\weights(\psi)=V_\weights(\phi)$. So, using the induction hypothesis,     
    $V_\weights(\inter{\psi}) \subseteq V_\weights(\psi)$. Moreover, 
    $\#\phi=\#\psi$, and we conclude by induction hypothesis: $|V_\weights(\phi)|\leq |\weights|^{\#\phi}$.
    \item Let $\psi=\phi_1 \U \phi_2$. By the semantics of
    $\U$, every satisfaction value of $\psi$ is a $\sup$ of a $\inf$ on
    the satisfaction values of $\phi_1$ and $\phi_2$ over all suffixes of a
    trace. But by induction hypothesis, the set of possible values in these $\sup$ and $\inf$ is finite, and thus these are $\max$ and $\min$. 
    Therefore, $V_\weights(\inter{\psi}) \subseteq V_\weights(\inter{\phi_1}) \cup V_\weights(\inter{\phi_2})$. By induction hypothesis, 
     $V_\weights(\inter{\phi_1}) \cup V_\weights(\inter{\phi_2}) \subseteq V_\weights(\phi_1) \cup V_\weights(\phi_2)= V_\weights(\psi)$. 
    Moreover, by the induction hypothesis $|V_\weights(\psi)| \leq |V_\weights(\phi_1)|+|V_\weights(\phi_2)| \leq
    |\weights|^{\#\phi_1}+|\weights|^{\#\phi_2} 
    \leq |\weights|^{\#\phi_1+\#\phi_2} = |\weights|^{\#\phi}$. 
    \item Let $\psi=\forall\pi\,\psi'$. As every satisfaction value of $\psi$ is
    an infimum on the satisfaction values of $\psi'$, we have $V_\weights(\inter{\psi}) \subseteq V_\weights(\inter{\psi'})$
    and thus a minimum by induction hypothesis, it holds that $V_\weights(\inter{\psi})
    \subseteq V_\weights(\psi)$ and further, by the induction hypothesis $|V_\weights(\psi)| = 
    |V_\weights(\psi')|\leq |\weights|^{\#\psi'}=|\weights|^{\#\psi}$. 
    \item Let $\psi=\exists\pi\, \psi'$. Similarly to the case above,
    every satisfaction value of $\psi$ is a maximum on the satisfaction values of
    $\psi'$. Thus, $V_\weights(\inter{\psi}) \subseteq V_\weights(\inter{\psi'})$ and then by induction hypothesis, $V_\weights(\inter{\psi}) \subseteq V_\weights(\psi)$ 
     and $V_\weights(\psi) = V_\weights(\psi')\leq |\weights|^{\#\psi'}=|\weights|^{\#\psi}$.\qedhere
\end{itemize}
\end{proof}

The exponential bound is tight, since it is already so for \LTLprop
\cite{Almagor14}. In general the inclusion $V_\weights(\inter\psi) \subseteq
V_\weights(\psi)$ is strict, as shown in this example:

\begin{example}\label{exmpl:bound satisfaction values} 
Let $\psi'=\forall \pi'\, (p_\pi \oplus_{1/2} p_{\pi'})$ and $\psi=\exists \pi\,\psi'$. We claim
that $0\in V_{\{0,1\}}(\inter{\psi'})$ (and thus $0\in V_{\{0,1\}}(\psi') = V_{\{0,1\}}(\psi)$) but $0\notin V_{\{0,1\}}(\inter\psi)$, so in particular $V_{\{0,1\}}(\inter\psi) \subsetneq V_{\{0,1\}}(\psi)$. First we show that $0 \in V_{\{0,1\}}(\inter{\psi'})$. Let $T=\{t\}$ with $t[0](p)=0$ and
$\Pi$ such that $\Pi(\pi)[0](p)=1$. Then
\begin{align*}
\hsem{\psi'}\Pi T & = \min_{t\in T}\hsem{p_\pi\oplus_{1/2}  p_{\pi'}} {\Pi[\pi' \mapsto t]} T = \hsem{p_\pi\oplus_{1/2}  p_{\pi'}}{\Pi[\pi' \mapsto t]} T\\
& = \frac{1}{2}(1-\hsem {p_\pi} {\Pi[\pi' \mapsto t]} T)+\frac{1}{2}\hsem {p_{\pi'}} {\Pi[\pi' \mapsto t]} T =0
\end{align*}
Next we show that for every trace assignment $\Pi$ and every set of
traces $T\subseteq \Traces_{\{0,1\}}$, $\hsem\psi \Pi T\geq\frac{1}{2}$.
Suppose there is a trace $t^*\in T$ such that $t^*[0](p)=0$. We observe then for
any $\Pi$ that $\hsem {p_\pi\oplus_{1/2} p_{\pi'}} {\Pi[\pi \mapsto t^*]} T\geq
\frac{1}{2}$ since the first addend of the evaluation is equal to
$\frac{1}{2}$ and further $\hsem{\psi'}{\Pi[\pi \mapsto t^*]} T=\min_{t\in
T}\hsem {p_\pi\oplus_{1/2} p_\pi'} {\Pi[\pi \mapsto t^*, \pi'\mapsto t]} T \geq
\frac{1}{2}$. Thus, $\hsem \psi \Pi T = \max_{t\in T}\hsem {\psi'} {\Pi[\pi \mapsto
t]} T\geq \hsem {\psi'} {\Pi[\pi \mapsto t^*]} T \geq \frac{1}{2}$.
In the same way, we obtain the result when there is a trace $t^*\in T$ such that
$t^*[0](p)=1$ since $\hsem {p_\pi \oplus_{1/2}  p_{\pi'}} {\Pi[\pi \mapsto t^*]} T\geq
\frac{1}{2}$ because of the second addend.\qed
\end{example}

Indeed, it is not surprising that the set $V_\weights(\psi)$ is a strict over-approximation of the possible satisfaction values of $\psi$ as the satisfiability for \HyperLTL (and thus of $\HyperLTLprop$) is undecidable \cite{DBLP:conf/concur/FinkbeinerH16}.  
Note that, as a consequence of this result, in the semantics of \HyperLTLprop, the various occurrences of operators $\inf$ and $\sup$ can be replaced by $\min$ and $\max$ respectively.

\subsection{\texorpdfstring{The model checking problem for \HyperLTLprop}{The model checking problem for HyperLTLprop}}
\label{sec:MC-HyperLTLprop}

The natural generalisation of the model checking of \HyperLTL 
to \HyperLTLprop is as follows

\begin{definition}
 \label{MCHyperLTL} Given a Kripke structure $\K$,
a closed \HyperLTLprop formula $\psi$, and a rational threshold $v\in[0,1]$, the model checking problem is to decide whether $\chsem\psi{\Traces(\K)}\geq v$ (or whether $\chsem\psi{\Traces(\K)}\leq v$, depending on the cases of interest). 
\end{definition}

We propose a model checking algorithm for  \HyperLTLprop. Its complexity will be non-elementary with a tower of 
exponentials depending  on the alternation depth of the \HyperLTL formula.
As a base case, it relies on the construction proposed in \cite{Almagor14} (Theorem 2.9 for the Boolean case, and Section 3.1 for the extension to weighted Kripke structures) 
to translate a formula of \LTLprop  for a Kripke structure $\K$ together with a set of targeted values into an NBA which accepts the traces of $\K$ whose semantics are in the target:

\begin{proposition}[\cite{Almagor14}]\label{prop:LTL[F]_to_NGBA} 
    Let $\phi$ be
    an \LTLprop formula, $\K$ be a Kripke structure,
    and $P\subseteq[0,1]$. There exists an NBA
    $\A^P_{\K, \phi}$ 
    such that for every trace $t\in \Traces(\K)$, it holds
    that $\sem \phi t\in P$ iff $\A^P_{\K, \phi}$ accepts $t$. Furthermore,
    $\A^P_{\K, \phi}$ has at most $|\weights|^{|\phi|^2} |\phi|$ states where $\weights$ is the set of weights in $\K$, and $|\phi|$ is the size of $\phi$.  
\end{proposition}
In this proposition, the set $P$ can take various forms, provided that testing
membership in $P$ of values in $[0,1]$ is decidable, so that the NBA
$\A^P_{\K, \phi}$ is computable. For instance, if $P$ is of the form $[0,
a)$ or $(a, 1]$, with some rational number $a$, the membership test can be
performed.

We extend this proposition to \HyperLTLprop formula by taking into account possible free variables as well as quantifications. We base the construction of such a resulting NBA on two crucial ideas used before separately. First, we make use of the NBA from Proposition~\ref{prop:LTL[F]_to_NGBA} on a Krikpe structure obtained as the synchronous product of multiple copies of $\K$ (to deal with the quantified variables of $\psi$), by treating the quantifier-free part of a \HyperLTLprop formula as an \LTLprop formula (as done in \cite{ClarkFink14} for \HyperLTL formulas with respect to \LTL and recall in 
Remark \ref{rem:HyperLTLtoLTL}).
Second, following the semantics of a quantified formula, we compute $\inf$ and $\sup$ over $\Traces(\K)$, by checking finitely many values, as $V_\weights(\psi)$ is finite (with $\weights$ the set of weights in $\K$).

%

For the first step, we extend Remark \ref{rem:HyperLTLtoLTL} to \HyperLTLprop,

\begin{remark}
\label{rem:HyperLTLproptoLTLprop}
Let $\psi = Q_1 \pi_1 \ldots Q_n \pi_n\, \phi$ be a closed \HyperLTLprop formula, $\phi$ being its quantifier-free part with $n$ free variables and 
$Q_i\in\{\exists,\forall\}$ for all $i\in\{1,\dots,n\}$. The formula $\phi$ turns out to be a formula of the logic \LTLprop defined over the set of atomic propositions $\AP_{\le n}$.
The semantics of \LTLprop formulas is given by a mapping from traces of $(\weights^{\AP_{\le n}})^\omega$ to $[0,1]$ 
such that $\sem \phi {t_\Pi} = \hsem \phi \Pi T$, for arbitrary $T$ (as the  \HyperLTLprop formula $\phi$ is non-quantified). 
\end{remark}

By an inductive generalisation of Proposition~\ref{prop:LTL[F]_to_NGBA} over the quantification prefix, one obtains: 

\begin{restatable}{proposition}{modelcheckinghyperLTLF}
    \label{thm:modelchecking_hyperLTL[F]} Let $\psi$ be a
    \HyperLTLprop formula, $P\subseteq[0,1]$, and $\K$ be a
    Kripke structure. One can build an NBA $\A_{\K,\psi}^{P}$ such that
    $\Lang(\A_{\K,\psi}^{P})$ is the set of traces $t_\Pi$ induced by $\Pi$, an assignment of free variables to traces of $\K$ verifying 
    $\chsem\psi{\TracesOf \K}(\Pi)\in P$. The size of $\A^P_{\K,\psi}$ is non-elementary, with the tower of exponentials of linear height in the alternation depth of $\psi$.
\end{restatable}

\begin{proof}
    Let $\psi = Q_1 \pi_1 \ldots Q_n \pi_n \, \phi$ be a closed \HyperLTLprop formula, $\phi$ being its quantifier-free part with $n$ free variables and 
    $Q_i\in\{\exists,\forall\}$ for all $i\in\{1,\dots,n\}$.
    
    By induction on the quantifier suffixes $\bar{Q}$ of $\psi$ such that $\bar{Q} \phi$ has $k$ free variables ($\bar{Q}$ is of length $n-k$), we show how to build, for all predicates $P\subseteq [0, 1]$, an NBA $\A_{\K,\bar{Q}\phi}^P$ over the alphabet $\weights^{\AP_{\le k}}$ for the formula $\bar{Q}\phi$ such that for all trace assignments $\Pi$, $t_\Pi \in\Lang(\A^P_{\K,\bar{Q}\phi})$ iff $\hsem{\bar{Q}\phi}{\Pi}{\TracesOf \K}\in P$ and $\Pi(\pi_i) \in \TracesOf \K$ for all $1 \le i \le k$.
    
    For the base case, relying on Remark \ref{rem:HyperLTLproptoLTLprop}, each quantifier-free \HyperLTLprop formula $\phi$ with $\pi_1, \ldots, \pi_n$ as free variables can be viewed as an \LTLprop formula over the set of atomic propositions $\AP_{\le n}$. Hence, appealing to Proposition~\ref{prop:LTL[F]_to_NGBA}, we can make use of the NBA $\A^P_{\K^n,\phi}$ of size exponential in the size of $\phi$. It fulfils, as expected, that for all trace assignments $\Pi$ over $\Traces$, $t_\Pi \in\Lang(\A^P_{\K^n,\phi})$ iff $\sem \phi {t_\Pi} \in P$ and $t_\Pi \in \TracesOf {\K^n}$
    iff $\hsem{\phi}{\Pi}{\TracesOf \K}\in P$ and $\Pi(\pi_i) \in \Traces(\K)$ for all $1 \le i \le n$ (by Lemma \ref{lem:traceproduct})\footnote{Note that the set $\TracesOf \K$ used in $\hsem{\phi}{\Pi}{\TracesOf \K}$ is chosen arbitrarily as the formula $\phi$ contains no quantifiers to evaluate.}.
    
    For the induction step, let us assume that the property holds for a formula $\psi' = \bar{Q}\phi$ and consider the next block of identical quantifiers in $\psi$.
    \begin{itemize}
      \item If the next block of quantifiers is existential, let $\psi'' := \exists \pi_{k+1} \ldots \exists \pi_{\ell}\, \psi'$ the subformula of $\psi$ with $k$ free variables $\pi_1, \ldots, \pi_k$. Notice that $\psi'$ has $\ell>k$ free variables $\pi_1, \ldots, \pi_k, \pi_{k+1}, \ldots, \pi_{\ell}$, and we can thus apply the induction hypothesis on it. Then, for each $c\in P \cap V_\weights(\psi'')$, we define the NBA $\A^{\{c\}}_{\K, \psi''}$ as
     \[\A^{\{c\}}_{\K, \psi'} \big | _{1 \ldots k} \cap  \A_{\K^k} 
     \cap 
     \overline{\A^{(c,1] \cap V_\weights(\psi')}_{\K, \psi'} \big | _{1 \ldots k} \cap  \A_{\K^k}}
     \]
     where $\overline{\mathcal B}$ denotes the complementation of the NBA $\mathcal B$, while the projection $|_{1 \ldots k}$ corresponds to the application on the transitions of the alphabet morphism mapping an element $S\in \weights^{\AP_{\le \ell}}$ to an element of $\weights^{\AP_{\le k}}$ (by only keeping the relevant atomic propositions).
     Then, we let 
     $\A^{P}_{\K, \psi''} = \bigcup_{c \in P \cap V_\weights(\psi'')} \A^{\{c\}}_{\K, \psi''}$. 
    
     Let us consider a trace assignment $\Pi$. Then, we have $t_\Pi \in \Lang(\A^{P}_{\K, \psi''})$ iff (by definition) there exists 
     $c\in P \cap V_\weights(\psi'')$ such that 
     $t_\Pi \in \Lang( \A^{\{c\}}_{\K, \psi'} \big | _{1 \ldots k})$, $t_\Pi \in  \Lang(\A_{\K,k})$  and
     $t_\Pi \notin \Lang(\A^{(c,1] \cap V_\weights(\psi')}_{\K, \psi'} \big | _{1 \ldots k} \cap  \A_{\K,k})$.
     By definition of projection, this is equivalent to
     \begin{equation}
     \forall 1\leq i\leq k \quad \Pi(\pi_i) \in \Traces(\K)\,,
        \tag{$\star$}\label{eq:tracesofK}
     \end{equation} 
     the existence of $c\in P \cap V_\weights(\psi'')$ and traces 
     $t_{k+1}, \ldots , t_{\ell}$ such that $t_{\Pi[\pi_{k+1} \mapsto t_{k+1},\ldots, \pi_{\ell} \mapsto 
     t_{\ell}]} \in \Lang(\A^{\{c\}}_{\K, \psi'})$ and the absence of traces $t'_{k}, \ldots t'_{\ell}\in \TracesOf \K$ such that $t_{\Pi[\pi_{k+1} \mapsto t'_{k+1},\ldots \pi_{\ell} \mapsto 
     t'_{\ell}]} \in \Lang(\A^{(c,1] \cap V_\weights(\psi')}_{\K, \psi'})$. By induction hypothesis,
     this is equivalent to \eqref{eq:tracesofK}, the existence of $c\in P \cap V_\weights(\psi'')$ and traces 
     $t_{k+1}, \ldots , t_{\ell}\in\TracesOf \K$ such that 
     $\hsem {\psi'}{\Pi[\pi_{k+1} \mapsto t_{k+1},\ldots \pi_{\ell} \mapsto  t_{\ell}]}{\TracesOf \K} = c$
     and the absence of traces $t'_{k+1}, \ldots, t'_{\ell}\in\TracesOf \K$ such that $\hsem {\psi'}{\Pi[\pi_{k+1} \mapsto t'_{k+1},\ldots \pi_{\ell} \mapsto  t'_{\ell}]}{\TracesOf \K} \in (c,1]$. 
     This is equivalent to say that \eqref{eq:tracesofK} and 
     \[\max _{t_{k+1}, \ldots, t_{\ell} \in \TracesOf \K} \hsem {\psi'}{\Pi[\pi_{k+1} \mapsto t_{k+1},\ldots \pi_{\ell} \mapsto  t_{\ell}]}{\TracesOf \K} \in P\]
     and thus, that $\hsem {\psi''}{\Pi}{\TracesOf \K} \in P$.
     
     \item If the next block of quantifiers is universal, let $\psi'' := \forall \pi_{k+1} \ldots \forall \pi_{\ell}\, \psi'$ the subformula of $\psi$ with $k$ free variables $\pi_1, \ldots, \pi_k$, we define the NBA $\A^{\{c\}}_{\K, \psi''}$ as
     \[\A^{\{c\}}_{\K, \psi'} \big | _{1 \ldots k} \cap  \A_{\K,k} 
     \cap 
     \overline{ \A^{[0,c) \cap V_\weights(\psi)}_{\K, \psi'} \big | _{1 \ldots k} \cap  \A_{\K,k}}
     \]
      and we let 
     $\A^{P}_{\K, \psi''} = \bigcup_{c \in P \cap V_\weights(\psi'')} \A^{\{c\}}_{\K, \psi''}$. The rest of the proof is as in the previous~case. 
     \end{itemize}  
\end{proof}

Note that for a closed formula $\psi$ and thus, the empty trace assignment, the NBA $\A^{P}_{\K,\psi}$ can accept at most the unique trace $t_{[ \, ]}$ defined as $u^\omega$ with $u$ the mapping of empty domain. 
As a corollary, for a closed \HyperLTLprop formula $\psi$, we obtain: 

\begin{theorem}
    The model checking problem of \HyperLTLprop is decidable with a
    non-elementary complexity.
    \begin{proof}
From Definition \ref{MCHyperLTL}, by Proposition \ref{thm:modelchecking_hyperLTL[F]}, it suffices to test the emptiness of the NBA $\A^{[0,a)}_{\K,\psi}$
(or of $\A^{(a,1]}_{\K,\psi}$, depending on the cases of interest). The complexity originates simply in the size of the constructed NBA.
\end{proof}
\end{theorem}

Note that the height of the tower of exponentials in the non-elementary complexity depends only on the alternation depth, similarly to the case of \HyperLTL~\cite{ClarkFink14}. Notice, as a final corollary, that we also obtain the decidability of the model checking problem for \HyperLTL over weighted Kripke structures (which was not known so far, to the best of our knowledge), since \HyperLTL is a fragment of \HyperLTLprop: 

\begin{corollary}
    The model checking problem of \HyperLTL (over weighted Kripke structures) is decidable with a non-elementary complexity.
\end{corollary}

\subsection{\texorpdfstring{Expressiveness of \HyperLTLprop in the Boolean case}{Expressiveness of HyperLTLprop in the Boolean case}}
\label{sec:HLTL[F]_to_HLTL}

In this section, we focus on Boolean Kripke structures, and 
we show that any \HyperLTLprop formula can be translated into a
somewhat equivalent \HyperLTL formula. This is an extension of the
analogous result for \LTLprop and \LTL \cite{Almagor14}. This implies \HyperLTLprop is not more expressive than \HyperLTL. 
However, it makes it a lot easier to express quantitative specifications, since the resulting \HyperLTL
formulas get very large and not very intuitive. 

\begin{restatable}{theorem}{thmHLTLFtoHLTL}
    \label{thm:HLTL[F]_to_HLTL} For every $P\subseteq [0,1]$ and
\HyperLTLprop formula $\psi$, there exists a \HyperLTL formula $\Bool(\psi,P)$ such that for all subset $T\subseteq \Traces_{\{0,1\}}$ and trace assignments $\Pi$, 
$$\hsem \psi \Pi T \in P  \text{~iff~} \hsem {\Bool(\psi,P)} \Pi T = 1$$
Moreover, $\Bool(\psi, P)$ has size at most $O(2^{(k+2)|\phi|})$ if $k$ is the quantifier depth of $\psi$, and $\phi$ is the quantifier-free part of $\psi$.
\end{restatable}

The proof of this theorem heavily uses the fact that the formula $\psi$ has only finitely many possible satisfaction values and thus, that only those which also belong to $P$ matter. Hence, as a first step, it is possible to consider $P$ by means of each of its elements $c$ that is also a possible satisfaction value of $\psi$. Then, recursively it is possible to define $\Bool(\psi,\{c\})$ as a Boolean combination of some $\Bool(\varphi_i,P_i)$ where $\varphi_i$ is a subformula of $\psi$ and $P_i$ is a subset of $P$ defined relatively to $c$.
Before giving the complete proof for Theorem~\ref{thm:HLTL[F]_to_HLTL}, a technical lemma has to be proved: we show that allowing the quantifiers and the Boolean operators (but not the temporal operators) in \HyperLTL to alternate does not change the expressiveness of the logic.

\begin{definition}\label{def:B(HyperLTL)}
We define the \emph{positive Boolean closure} $\mathcal{B}^+(\HyperLTL)$ of \HyperLTL by the following grammar:
\begin{align*}
            \psi &::=  \exists\pi\, \psi \mid \forall\pi\, \psi \mid \psi \vee \psi \mid \psi\wedge\psi \mid \phi \\
            \phi &::=  p_\pi \mid \lnot\phi \mid \phi \vee \phi \mid \X\phi \mid \phi\U \phi
\end{align*}
for all atomic propositions $p\in AP$ and variables $\pi\in\mathcal{V}$.
\end{definition}

The semantics of formulas is extended in the natural way:
\[\hsem{\psi_1 \vee \psi_2}{\Pi}{T} = \max(\hsem{\psi_1}{\Pi}{T}, \hsem{\psi_2}{\Pi}{T} ) \qquad 
 \hsem{\psi_1 \wedge \psi_2}{\Pi}{T} = \min(\hsem{\psi_1}{\Pi}{T}, \hsem{\psi_2}{\Pi}{T} ) \]
\HyperLTL is exactly the fragment of $\mathcal{B}^+$(\HyperLTL) formula in \emph{prenex form} (i.e.~that begins by a sequence of quantifiers, followed by a quantifier-free part).

    The notion of alternation depth $\alpha(\psi)$ is extended to formulas from $\mathcal{B}^+$(\HyperLTL)  by letting 
    $$  \alpha_Q(\psi_0\circ\psi_1)  = \max\{\alpha_Q(\psi_0),\alpha_Q(\psi_1)\} \ \text{ where } \circ\in\{\wedge,\vee\} \quad \text{for } Q\in\{\forall,\exists\}$$
    and recalling that 
      \begin{align*}
       \alpha_Q(\psi_0) & = 0 \ \text{ if } \psi_0 \text{ is quantifier-free }\\
        \alpha_Q(Q\pi\, \psi_0) & = \alpha_Q(\psi_0) \\
        \alpha_Q(\bar{Q}\pi\, \psi) & = \alpha_{\bar{Q}}(\psi_0)+1 \ \text{ where } \bar{Q} \text{ is the quantifier dual of } Q\\
    \end{align*}

    One has the following folklore result:
\begin{proposition}\label{prop:altern_depth}
    Let $\psi$ be a $\mathcal{B}^+(\HyperLTL)$ formula. Then there exists an equivalent formula $\psi^*$ in \HyperLTL with alternation depth $\alpha(\psi^*)\leq\alpha(\psi)+1$ and the same number of existential and universal quantifiers.
\end{proposition}
\begin{proof}
%
    In the following we will simplify notation and just write $\exists^*$ for a series of existential quantified variables, $\exists^+$ for a non-empty series of existential quantified variables and $\forall^*$, or $\forall^+$ respectively, for series of universal quantified variables. Since we are allowed to rename bound variables, we can assume that each variable occurs only once.

    We begin with a basic observation. If $\psi$ is a formula in prenex form with $\alpha_\exists(\psi)=k-1$, for a $k>0$, it has to be of shape $\psi=\exists_0^*\forall_1^+\exists_2^+\dots\bar{Q}_{k-1}^+Q_k^+\psi'$ for a quantifier-free $\psi'$, where $Q_k$ depends on the parity of $k$, i.e. $Q=\exists$ iff $k$ is even. Note that this implies in particular $\alpha(\psi)\leq\alpha_\exists(\psi)\leq\alpha(\psi)+1$ for a formula in prenex form. We will use this fact in our conclusion.
    
    The next step is to show by induction over $\psi$ that for each $Q\in\{\forall,\exists\}$ there is a formula $\psi^Q\equiv\psi$ in prenex form which is stable regarding $\alpha_Q$ in the sense that $\alpha_Q(\psi^Q)=\alpha_Q(\psi)$, and in addition it contains the same number of each quantifier as $\psi$.
    \begin{itemize}
        \item Let $\psi=\exists\pi.\psi_0$. By hypothesis there is a formula in prenex form $\psi_0^{\exists}\equiv\psi_0$, containing the same number of each quantifier and satisfying $\alpha_\exists(\psi_0^{\exists})=\alpha_{\exists}(\psi_0)$. Thus, $\psi\equiv\exists\pi.\psi^{\exists}_0$ and the latter is in prenex form conserving the number of each quantifier.
        Then we obtain by definition $\alpha_{\exists}(\psi)=\alpha_{\exists}(\exists\pi.\psi_0)=\alpha_{\exists}(\psi')$ and by hypothesis also  $\alpha_{\exists}(\exists\pi.\psi_0^{\exists})=\alpha_{\exists}(\psi_0^{\exists})=\alpha_{\exists}(\psi_0)$ for the existential measure.
        And moreover for our universal measure, by definition $\alpha_{\forall}(\exists\pi.\psi_0)=\alpha_{\exists}(\psi_0)+1$ and by hypothesis $\alpha_{\forall}(\exists\pi.\psi^{\exists}_0)=\alpha_{\exists}(\psi^{\exists}_0)+1=\alpha_{\exists}(\psi_0)+1$.
        \item The case $\psi=\forall\pi.\psi_0$ is analogous.
        \item Let $\psi=\psi_0\circ\psi_1$, where $\circ\in\{\vee,\wedge\}$. We will begin by constructing the prenex formula which preserves $\alpha_\exists$. 
        \item 
        By hypothesis there exist $\psi_i^\exists\equiv\psi_i$ for $i\in\{0,1\}$ in prenex form, containing the same number of each quantifier and satisfying $\alpha_{\exists}(\psi_i)=\alpha_{\exists}(\psi_i^\exists)=k_i-1$ for some $k_i>0$ and w.l.o.g. $k_0\geq k_1$. By our remark above we can assume w.l.o.g. that the formulas are of the shapes 
        \begin{align*}
        \psi_1^\exists & =\exists_{0,0}^*\forall_{0,1}^+\exists_{0,2}^+\dots Q_{0,k_0}^+\, \psi'_0 \quad \text{and}\\
        \psi_2^\exists & = \exists_{1,0}^*\forall_{1,1}^+\exists_{1,2}^+\dots Q_{1,k_1}^+\, \psi'_1,
        \end{align*}
        where each $\psi'_i$ is quantifier-free and each $Q_{i,k_i}\in\{\forall,\exists\}$ depends on the parity of $k$. Since we are allowed to rename bound variables, we can assume that we are not quantifying over variables which appear in the other formula. This allows us to put the quantifiers at the beginning, and we obtain:
        \begin{align*}
            \psi & = \psi_0 \circ \psi_1 \\
            & \equiv \psi_0^\exists \circ \psi_1^\exists \\
            & = (\exists_{0,0}^*\forall_{0,1}^+\dots Q_{0,k_0}^+\, \psi'_0) \circ (\exists_{1,0}^*\forall_{1,1}^+\dots Q_{1,k_1}^+\, \psi'_1) \\
            & \equiv \exists_{0,0}^*\exists_{1,0}^*.(\forall_{0,1}^+\exists_{0,2}^+\dots Q_{0,k_0}^+\, \psi'_0 \circ \forall_{1,1}^+\exists_{1,2}^+\dots Q_{1,k_1}^+\, \psi'_1) \\
            & \equiv \dots \\
            & \equiv \exists_{0,0}^*\exists_{1,0}^*\forall_{0,1}^+\forall_{1,1}^+\dots Q_{0,k_1}^+Q_{1,k_1}^+Q_{0,k_1+1}^+\dots Q_{0,k_0}^+\, (\psi'_0 \circ \psi'_1)
        \end{align*}
        Let us call the formula we obtained in the last line $\psi^\exists$. By
        the definition of $\alpha_\exists$ it holds that
        $\alpha(\psi^\exists)=k_0$ and on the other hand by definition and our
        hypothesis:
        \[\alpha_\exists(\psi)=\max\{\alpha_\exists(\psi_0),\alpha_\exists(\psi_0)\}=\max\{\alpha_\exists(\psi_0^\exists),\alpha_\exists(\psi_0^\exists)\}=\alpha_\exists(\psi_0^\exists)=k_0\]
        In particular the construction preserves the number of existential and
        universal quantifiers.
        \item The symmetrical procedure allows us to construct $\psi^\forall$ satisfying $\alpha_\forall(\psi^\forall)=\alpha_\forall(\psi)$ from $\psi_0^\forall$ and $\psi_1^\forall$ which exist by induction hypothesis.

        We can now conclude the assertion by choosing $\psi^*=\psi^\exists$ (we could also take $\psi^\forall$) and the fact that for every $\psi'$ in prenex form it holds that $\alpha(\psi')\leq\alpha_\exists(\psi')\leq\alpha(\psi')+1$, mentioned above.  Note that the proof also provides an algorithm to construct the prenex form, controlling the number of quantifier alternations. \qedhere
    \end{itemize}
\end{proof}

\begin{proof}[Proof of Theorem~\ref{thm:HLTL[F]_to_HLTL}]
    We construct the formula $\Bool(\psi,P)$ inductively on $\psi$. The correctness follows by
induction, which we only prove in detail for the quantified case, which is new compared to~\cite{Almagor14}.
\begin{itemize}
    \item If $\psi=p_\pi$. Then $\Bool(\psi,P)$ is equal to the formula $\true$ if $0 \in P\text{ and }1\in P$, $p_\pi$ if $0 \notin P\text{ and }1 \in P$, 
           $\lnot p_\pi$ if  $0 \in P\text{ and }1\notin P$ and $\false$ 
           if $0 \notin P \text{ and }1 \notin P$. 

    \item If $\psi=f(\phi_1,\dots,\phi_k)$ with $f\in \funs$, then we let 
        \[
        \Bool(\psi,P)
        :=\bigvee_{\substack{d_1 \in V_{\{0,1\}}(\phi_1),\ldots,d_k \in V_{\{0,1\}}(\phi_k),\\
        f(d_1,\dots,d_k)\in P}}\Bool(\phi_1,\{d_1\})\land \dots \land \Bool(\phi_k,\{d_k\}) \]
    \item If $\psi=\X\phi$, then we let
    $\Bool(\psi,P):=\X \Bool(\phi,P)$. 
    \item If $\psi=\phi_1 \U \phi_2$, then we let $\Bool(\phi_1 \U
    \phi_2):=\bigvee_{c\in P\cap V_{\{0,1\}}(\psi)}\Bool(\phi_1 \U \phi_2,\{c\})
    $
    where
    \[
    \Bool(\phi_1 \U \phi_2,\{c\}):= \Bool(\phi_1,[c,1])\U \Bool(\phi_2,[c,1]) \land\lnot (\Bool(\phi_1,(c,1])\U \Bool(\phi_2,(c,1]))     
    \]
    This is motivated by the semantics of the $\U$ operator as a supremum/maximum operator. Intuitively, the first term ensures that the semantics is greater or equal to $c$. The second term ensures that the semantics is not strictly greater.

    \item Let $\psi=\forall \pi\,\psi'$. We start by considering the case
    $P=\{c\}$. By the semantics of the $\forall$ operator it holds that
    $\hsem{\forall\pi\, \psi'} \Pi T = c$ is equivalent to $\min_{t\in T}\hsem{\psi'} {\Pi[\pi \mapsto t]} T=c$, and thus, obviously to 
    $ \exists t \in T \; \hsem{\psi'}{\Pi[\pi \mapsto t]} T=c  \text{ and } \forall t'\in T\; \hsem{\psi'}{\Pi[\pi \mapsto t']} T\geq c$. 
    By induction hypothesis, this is equivalent to $\exists t\in T \; \hsem{\Bool(\psi',\{c\})}{\Pi[\pi\mapsto t]} T = 1 \text{ and } \forall t'\in T\; 
    \hsem{\Bool(\psi',[c,1])}{\Pi[\pi\mapsto t']} T  = 1$. This is equivalent by the semantics of \HyperLTL to 
    $\hsem{\exists\pi \, \Bool(\psi',\{c\})\land \forall\pi\, \Bool(\psi',[c,1])}{\Pi} T = 1$. 
    
    For a general $P\subseteq [0,1]$ we add a disjunction over all possible
    values in $P$ that can be a possible satisfaction value for $\psi$
    \begin{equation}\label{eq:1}
    \bigvee_{c\in P \cap V_{\{0,1\}}(\psi)}\exists \pi\, \Bool(\psi',\{c\})\land \forall\pi\, \Bool(\psi',[c,1])   
    \end{equation}
     This is a priori not a \HyperLTL formula, but it is in the
     positive Boolean closure of $\HyperLTL$. Thus, we can appeal to Proposition \ref{prop:altern_depth}  to obtain an equivalent formula 
     $\Bool(\forall\pi\, \psi',P)$.

    \item Let $\psi=\exists \pi\, \psi'$. Then we define, with an analogous
    reasoning, $\Bool(\psi,P)$ as the \HyperLTL equivalent of the formula 
    $\bigvee_{c\in P \cap V_{\{0,1\}}(\psi)}\exists \pi\, \Bool(\psi',\{c\})\land \forall \pi\, \Bool(\psi',[0,c])$. 
\end{itemize}

With respect to the size of the formula $\Bool(\psi, P)$, the size of the large disjunction in equation (\ref{eq:1}) can grow exponentially, as shown in~Lemma~\ref{lem:number_satisfac_values}, which directly implies that $|\Bool(\psi,P)|\in O(2^{(k+2)|\phi|})$ if $k$ is the quantifier depth of $\phi$, and $\phi$ is the quantifier-free part of $\psi$.
\end{proof}

Some Boolean functions can only be expressed by exponential size Boolean
formulas~\cite{Shannon49}. Therefore, as noticed in \cite{Almagor14} already,
the translation of an \LTLprop formula to an equivalent \LTL formula can grow
exponentially: in our setting, there exists a sequence of quantifier-free
formulas $\phi$, of growing length, and a predicate $P\subseteq[0,1]$ such that
$|\Bool(\phi,P)|\in \Omega(2^{|\phi|})$. In \cite[Theorem 2.8]{Almagor14}, 
again in the case of \LTLprop, a thorough discussion on the complexity of
computing $\Bool(\phi, P)$ explains that the computation can be performed in
PSPACE (under reasonable assumptions  on the complexity of the functions $f$
that appear in the formula), and that a super-polynomial blow-up (i.e.~at least
$|\varphi|^c$ for all constants $c$) is unavoidable.

Finally, note that the above translation of a \HyperLTLprop formula $\psi$ into an equivalent Boolean
\HyperLTL formula (whose model checking is non-elementary~\cite{ClarkFink14}) yields
an alternative model checking algorithm for \HyperLTLprop, with non-elementary complexity. However, the height of exponentials depends on the quantifier depth of $\psi$ (instead on the 
alternation depth in Theorem~\ref{thm:modelchecking_hyperLTL[F]}); indeed, the alternation depth of $\Bool(\psi, P)$ depends on the quantifier depth of the original formula $\psi$ (and not 
only on its alternation depth).

\section{Temporal quality for \HyperLTL}\label{sec:hyperLTL_disc[D]} 
\label{sec:tempHyperLTL}

\subsection{Definitions and examples}

In this section we consider another qualitative extension of \HyperLTL named
\HyperLTLtemp (for \HyperLTL \emph{with temporal quality}). This logic is based
upon  \LTLtemp, another qualitative extension of LTL proposed in
\cite{Almagor14}\footnote{This logic is called $\LTL^\mathit{Disc}[\discfuns]$
in that paper.}. This latter generalises \LTL by considering new temporal
operators $\U_\eta$ where $\eta$ is taken from a set $\discfuns$ of discounting
sequences which parameterises the logic: $(\eta_i)_{i\in \N}$ is a
\emph{discounting sequence} if for all $i$, $\eta_i\in [0,1]$,
$\lim_{i\to\infty}\eta_i=0$, and $\eta$ is decreasing. Note, that this implies
in particular that $\eta_i>0$ for all $i\in\N$. Common examples for discounting
sequences are $\eta_i=\alpha^i$, for some $\alpha\in(0,1)$, and
$\eta_i=\frac{1}{i+1}$.

In contrast to propositional quality which allows mitigating the weights of  subformulas, temporal quality aims to give weights depending on \emph{when} events happen. Intuitively, for a formula $\phi_1 \U_\eta \phi_2$, the further $\phi_2$ is considered the least its value will impact the value of $\phi_1 \U_\eta \phi_2$.

\begin{definition} \label{def:HyperLTL[D]} 
    Assuming $\discfuns$ is a finite set of discounting sequences, the syntax of \HyperLTLtemp  is given by 
    \begin{align*}
    \psi & ::= \forall\pi\, \psi \mid \exists\pi\, \psi \mid \phi  \\
    \phi & ::=  p_\pi \mid \neg \phi \mid \phi \vee \phi \mid \X\phi \mid \phi \U \phi \mid \phi \U_{\eta} \phi
\end{align*}
for all $p\in \AP$, $\pi\in \mathcal V$,
 and $\eta\in\discfuns$. The \HyperLTL semantics is extended for 
    \HyperLTLtemp~as 
    \[\hsem{\phi_1 \U_{\eta} \phi_2} \Pi T  = \sup_{i\geq0}\Big(\min\big(\eta_i \, \hsem{\phi_2}{\Pi[i,\infty]}T,\min_{0\leq j<i}(\eta_j \, \hsem{\phi_1}{\Pi[j,\infty]}T)\big)\Big)\]
    for all trace assignments $\Pi$ over $\Traces$, and a non-empty set of traces $T \subseteq \Traces$. 
\end{definition}

We add as syntactic sugar the operators $\F_\eta\phi=\true\U_\eta\phi$ and
$\G_\eta\phi=\lnot\F_\eta\lnot\phi$. Their semantics can be simplified as 
\[\hsem{\F_\eta\phi} \Pi T
=\sup_{i\geq0}(\eta_i \, \hsem \phi {\Pi[i,\infty]}T)$ \text{ and }
$\hsem{\G_\eta\phi}\Pi T=\inf_{i\geq0}(1-\eta_i \, (1-\hsem
\phi{\Pi[i,\infty]}T))\]

We propose now two examples showing this extension of \HyperLTL with temporal quality allows to model hyperproperties; again it has a very different flavour than other quantitative extensions of \HyperLTL as the ones from \cite{DBLP:journals/corr/abs-1905-13514}, for instance:

\begin{example}\label{ex:OD-disc}
    We can formulate another version of observational determinism
    in the logic \HyperLTLtemp. We refine $\varphi_{\mathsf{OD}}$ by replacing the global operator with a discounted one: 
    \begin{equation}
    \varphi_{\mathsf{OD}}^{\mathrm{temp}}  \quad = \quad  
    \forall\pi\, \forall\pi' \quad \underbrace{\lowequiv \pi {\pi'} \ \rightarrow \ \G_\eta \lowequiv \pi {\pi'}}_{\phi'}  
\end{equation}
    Note that, since discounted sequences are injective, the subformula $\phi'$
    gives  
    information about the moment where the two traces $\pi$ and $\pi'$ disagree on low atomic propositions. 
    The prefix of universal quantifiers captures this earliest moment in the full system.
    This could be of practical interest since the further in the future we leak
    information about secrets, the less probable a malicious
    adversary will dedicate that much time on an attack. So this amounts to verify
    that the satisfaction value of $\varphi_{\mathsf{OD}}^{\mathrm{temp}} $ is greater than a certain threshold.\qed
\end{example}

\begin{example}\label{ex:noninference-disc} Another version of non-inference in \HyperLTLtemp can also be reformulated. We refine the formula $\varphi_{\mathsf{NI}}$ by discounting the second part of the formula on the equality of low atomic propositions: 
    \begin{equation}
     \varphi_{\mathsf{NI}}^{\mathrm{temp}}  \quad = \quad 
        \forall\pi\, \exists\pi'\quad \G\dummy{\pi'}\land \G_\eta \lowequiv \pi {\pi'} 
    \end{equation}
    For a Boolean Kripke structure $\K$, it computes either $1$ if the
    non-inference is satisfied, or $1-\eta_i$ where $i$ corresponds to
    the greatest value where every possible trace $ t \in \Traces(\K)$ has a
    trace $t' \in \Traces(\K)$ satisfying $\G \dummy{t'}$ and $\lowequiv t
    {t'}$ until step $i$. Hence, it computes a value representing the longest
    possible duration where the non-inference is satisfied in~$\K$.\qed
\end{example}

\subsection{\texorpdfstring{Model checking for \HyperLTLtemp}{Model checking for HyperLTLtemp}}

The model checking for \HyperLTLtemp aims to compare with some threshold the semantical value of a formula when applied to the set of traces of some Kripke structure.  

For computability reasons, we assume from now on that the discounting sequences
$(\eta_i)_{i\in \N}$ are recursively enumerable sequences of rational numbers. Being recursively enumerable and strictly decreasing, the elements of such a sequence form therefore a recursive set of rationals.

Notice that, even in the case of Boolean Kripke structures, there is no hope to apply the strategy used in Section~\ref{sec:HLTL[F]_to_HLTL} to model check \HyperLTLprop by translating a formula and a threshold to an equivalent Boolean \HyperLTL formula. Indeed, it is already impossible
to find some equivalent \LTL formula for any \LTLtemp formula, since \LTLtemp formulas can generate non $\omega$-regular languages, as noticed in \cite{Almagor14} and recalled in the example below. 

\begin{example}\label{ex:no-lasso}
    Consider for instance the formula $\phi = \G(\F_\eta p)\lor \F\G \lnot p$.
    For a trace $t\in \Traces_{\{0,1\}}$, either $\sem{\phi} t=1$ if $p$ is seen only finitely often
    in $t$ (with weight 1), or $\sem{\phi}t=\liminf_k \eta_{i_{k+1} - i_k-1}$
    where $(i_k)_{k\in \N}$ is the sequence of positions where $p$ holds in
    $t$. In particular, $\sem\phi t$ is always different from 0 when $t$ is a
    lasso trace; however $\sem\phi t$ could be equal to 0 for some non-lasso trace $t_\infty$,
    for instance when $i_k = k^2$. This shows that the language of the traces
    $t$ such that $\sem{\phi}t =0$ is not $\omega$-regular as it is non-empty,
    yet does not contain any lasso trace (although, there exist lassos with a
    semantics arbitrarily close to $0$).\qed
\end{example}

Nonetheless, the notion of lassos remains useful regarding model checking purposes, and we introduce the notion of \emph{lasso assignments}:
\begin{definition}\label{def:lassosassign}
A trace assignment $\Pi$ is a lasso 
assignment if for all trace variables $\pi$ for which it is defined, $\Pi(\pi) \in \Lassos$. Equivalently, the associated trace $t_\Pi$ is also a lasso.
\end{definition}

Over lasso assignments, we show that Büchi automata can be used to describe the set of traces satisfying some \HyperLTLtemp formula with respect to some threshold. This will also imply that, on a semantical point of view, formulas coincide when interpreted over traces and over lassos, and thus, will provide some automata-based tools to decide model checking for some fragments of \HyperLTLtemp.

\subparagraph*{Automata models for \HyperLTLtemp}

The following result from \cite{Almagor14} 
relates models of $\LTLtemp$ formulas in a Krikpe structure, and traces accepted by some NBA. 

  \begin{proposition}[\cite{Almagor14}]\label{prop:ltldisc_nba_precice} Given
        an $\LTLtemp$ formula $\phi$, a Kripke structure $\K$, and a threshold
        $v\in[0,1]$, 
        for an ordering relation $\bowtie$ in $\{<,>\}$,
        there exists an NBA $\A^{\bowtie v}_{\K,\phi}$ 
        such that for every trace $t\in \Traces$:
        \begin{enumerate}
            \item If $\sem\phi t \bowtie v$, then $\A^{ \bowtie v}_{\K,\phi}$  accepts $t$.
            \item If $\A^{ \bowtie v}_{\K,\phi}$ accepts $t$ and $t$ is a lasso, then $\sem \phi t  \bowtie v$.
        \end{enumerate}
    \end{proposition}

We build upon this result and extend it to formulas with quantifiers, and going from the classical semantics (in the first item) to the lasso semantics (in the second item) using 
Definition \ref{def:tPI} of trace assignment encoding and the following adaptation of Remark \ref{rem:HyperLTLtoLTL} to the case of \HyperLTLtemp. 

\begin{remark}
\label{rem:HyperLTLtemptoLTLtemp}
Let $\psi = Q_1 \pi_1 \ldots Q_n \pi_n \, \phi$ be a closed \HyperLTLtemp formula, $\phi$ being its quantifier-free part with $n$ free variables and 
$Q_i\in\{\exists,\forall\}$ for all $i\in\{1,\dots,n\}$. The formula $\phi$ turns out to be a formula of the logic \LTLtemp defined over the set of atomic propositions $\AP_{\le n}$.
The semantics of \LTLtemp formulas is given by a mapping from traces of $(\weights^{\AP_{\le n}})^\omega$ to $[0,1]$ 
such that $\sem \phi {t_\Pi} = \hsem \phi \Pi T$, for arbitrary $T$ (as the  \HyperLTLtemp formula $\phi$ is non-quantified). 
\end{remark}

\begin{restatable}{lemma}{ornabetter}
    \label{lem:orna++} Let $\bowtie$ be an ordering relation in $\{<,>\}$ and $\underline{\bowtie}$ its reflexive closure. Then, 
    for all formulas $\psi$ of \HyperLTLtemp, for all Kripke structures $\K$, for all
    $v\in [0,1]$, one can build an NBA $\A_{\K, \psi}^{\bowtie v}$  such that
    for every trace assignment~$\Pi$ over the free variables of $\psi$, it holds that:
    \begin{enumerate}
        \item\label{item:logic->auto} if $\hsem\psi\Pi {\Traces(\K)} \bowtie v$  and all traces in $\Pi$ belong to $\Traces(\K)$, then $\A_{\K, \psi}^{ \bowtie v}$ accepts $t_{\Pi}$;
        \item\label{item:auto->logic} if $\A_{\K,\psi}^{\bowtie v}$  accepts a lasso $t_\Pi$, then all traces in $\Pi$ belong to $\Lassos(\K)$ and $\hsem\psi \Pi {\Lassos(\K)} \ \underline{\bowtie} \ v$.
    \end{enumerate}
\end{restatable}
\begin{proof}

    The proof is by induction on the quantifier alternation of the formula $\psi$.
    
    If $\psi$ is non quantified, the result follows from Proposition~\ref{prop:ltldisc_nba_precice}, Remark~\ref{rem:HyperLTLtemptoLTLtemp}, and  Lemma~\ref{lem:traceproduct}.
    For item~2, we would even obtain a strict (and thus stronger) inequality. Notice that switching between the lasso semantics and the classical one does not make any difference for a non quantified formula.

    If $\psi = \exists \pi_1\ldots \pi_n\; \psi'$, we let $\A_{\K, \psi}^{>v}$ be obtained from $\A_{\K, \psi'}^{>v}$ by 
    projecting away its $\pi_1\ldots \pi_n$-components in the alphabet. Then, we prove the two items, for every trace assignment $\Pi$ 
    (over the free variables of $\psi$).
        \begin{enumerate}
            \item If $\hsem \psi \Pi {\Traces(\K)}>v$ and all traces in $\Pi$ belong to $\Traces(\K)$, by definition of the semantics of the existential quantifier as a supremum (and because the operator $>$ is strict),
            there exist traces $t_1, \ldots t_n \in \Traces(\K)$ such that $\hsem{\psi'}
            {\Pi[\pi_1 \mapsto t_1, \ldots, \pi_n \mapsto t_n]}{\Traces(\K)}>v$. By induction
            hypothesis, this implies that $t_{\Pi[\pi_1 \mapsto t_1, \ldots, \pi_n \mapsto t_n]}$ is accepted by
            $\A_{\K, \psi'}^{>v}$. This implies that $t_\Pi$ is accepted by $\A_{\K, \psi}^{>v}$.
            \item  Suppose that $\A_{\K, \psi}^{>v}$ accepts a lasso trace $t_\Pi$ and $\Pi$, the corresponding assignment. Consider the NBA $\mathcal B$ obtained from
            $\A_{\K, \psi'}^{>v}$ by fixing the components in the free variables of
            $\psi$ to the lasso $t_\Pi$ and letting the free variables $\pi_1,\ldots \pi_n $ take any traces from $\TracesOf \K$ ($\mathcal B$ is still a finite NBA since $\Pi$ is a lasso assignment). Then $t_\Pi$ is trivially accepted by the NBA obtained by projecting away the $\pi_1 \ldots \pi_n$-components in $\mathcal B$. Therefore, this projection is non-empty;
            this can be formally written as $\Lang(\A_{\K, \psi'}^{>v}) \cap (\{t_\Pi\}\times \Lang(\A_{\K^n})) 
            \neq \emptyset$, and the NBA $\A_{\K, \psi'}^{>v}$
            accepts a lasso of the form $t_{\Pi[\pi_1\mapsto t_1, \ldots, \pi_n \mapsto t_n]}$. In particular, $t_1, \ldots, t_n$ are lassos. By induction hypothesis, all traces in $\Pi[\pi_1\mapsto t_1, \ldots, \pi_n\mapsto t_n]$ belong to $\Lassos(\K)$, and $\hsem{\psi'}
            {\Pi[\pi_1 \mapsto t_1, \ldots, \pi_n \mapsto t_n]}{\Lassos(\K)} \geq v$. In particular, all traces of $\Pi$ belong to $\Lassos(\K)$. Moreover, by definition of the semantics of the existential quantifier as a supremum
            (and since the inequality is large), we
            obtain $\hsem\psi \Pi {\Lassos(\K)} \geq v$.
        \end{enumerate}

    If $\psi = \forall \pi_1 \ldots \pi_n\; \psi'$ with $\psi'$ having $k+n$ free variables, using the equivalence with the ``formula'' $\lnot \exists \pi\; \lnot \psi'$, we let $\A_{\K, \psi}^{>v}$ be equal to 
    $\overline{(\overline{\A_{\K, \psi'}^{>v}} \cap \A_{\K^{k+n}})\big|_{1..k}} \cap \A_{\K^{k}}$. Then, we
    prove the two items, for every trace assignment $\Pi$ (over the free
    variables of $\psi$). 
    \begin{enumerate}
        \item If $\hsem \psi \Pi {\Traces(\K)}>v$ and all traces in $\Pi$ belong to $\Traces(\K)$, by definition of the semantics of the universal quantifier as an infimum (and because the operator $>$ is strict), for
        all traces $t_1, \ldots t_n \in \Traces(\K)$, $\hsem{\psi'}{\Pi[\pi_1 \mapsto
        t_1, \ldots \pi_n \mapsto t_n]}{\Traces(\K)}>v$. By induction hypothesis, this implies that
        $t_{\Pi[\pi_1 \mapsto t_1, \ldots, \pi_n \mapsto t_n]}$ is accepted by $\A_{\K, \psi'}^{>v}$, for all $t_1, \ldots, t_n \in \Traces(\K)$.
        If $t_\Pi$ was not accepted by $\A_{\K, \psi}^{>v}$, since it only contains traces of $\Traces(\K)$, it would not be accepted by
        $\overline{(\overline{\A_{\K, \psi'}^{>v}} \cap \A_{\K^{k+n}})\big|_{1..k}}$, and thus would be accepted by $(\overline{\A_{\K, \psi'}^{>v}} \cap \A_{\K^{k+n}})\big|_{1..k}$.
        Thus, there would exist
        $t_1, \ldots, t_n \in \Traces(\K)$ such that $t_{\Pi[\pi_1 \mapsto t_1, \ldots, \pi_n \mapsto t_n]}$ is
        accepted by $\overline{\A_{\K, \psi'}^{>v}}$, and thus
        would not be accepted by $\A_{\K, \psi'}^{>v}$. This
        contradicts the previous assertion.
        \item  Suppose that $\A_{\K, \psi}^{> v}$ accepts $t_\Pi$ and $\Pi$ is a
        lasso assignment. Since $t_\Pi$ is accepted by $\A_{\K^{k}}$, all traces of $\Pi$ belong to $\Traces(\K)$, and thus to $\Lassos(\K)$. Moreover, $t_\Pi$ is not accepted by 
        $(\overline{\A_{\K, \psi'}^{>v}} \cap \A_{\K^{k+n}})\big|_{1..k}$. Thus, no
        traces $t_1, \ldots, t_n \in \Traces(\K)$ are such that $t_{\Pi[\pi_1 \mapsto t_1, \ldots, \pi_n \mapsto t_n ]}$ is
        accepted by $\overline{\A_{\K, \psi'}^{>v}}$. As a consequence,
        for all traces $t_1, \ldots, t_n  \in \Traces(\K)$, $t_{\Pi[\pi_1 \mapsto t_1, \ldots, \pi_n \mapsto t_n]}$ is
        accepted by $\A_{\K, \psi'}^{> v}$. This is in particularly true for all lassos $t_1, \ldots, t_n \in \Lassos(\K)$. By induction hypothesis, $\hsem{\psi'}{\Pi[\pi_1 \mapsto t_1, \ldots, \pi_n \mapsto t_n]}{\Lassos(\K)}\geq v$ for all $t\in \Lassos(\K)$.
        By definition of the universal quantifier as an infimum (and since the inequality is large), we obtain $\hsem \psi \Pi {\Lassos(\K)}\geq v$.
    \end{enumerate}

    If $\psi = \exists \pi_1, \ldots \pi_n \, \psi'$, the NBA $\A_{\K, \psi}^{<v}$ is obtained by noticing that for all $T\subseteq \Traces(\K)$, $\hsem{\exists\pi_1\ldots \pi_n \, \psi'}{\Pi}{T} = 1 - \hsem{\forall \pi_1\ldots \pi_n \, \lnot \psi'}{\Pi}{T}$. In particular, 
    $\hsem{\exists\pi_1 \ldots \pi_n\, \psi'}{\Pi}{T}<v$ if and only if $\hsem{\forall \pi_1\ldots \pi_n  \, \lnot \psi'}{\Pi}{T} > 1-v$. We thus let $\A_{\K, \psi}^{<v}$ be equal to $\A_{\K, \forall \pi \, \lnot\psi'}^{>1-v}$.

    Similarly, if $\psi = \forall \pi_1 \ldots \pi_n  \, \psi'$, we let $\A_{\K, \psi}^{<v}$ be equal to $\A_{\K, \exists \pi \, \lnot \psi'}^{>1-v}$.
\end{proof}

Taking only into account the dependency in the quantifier part of \HyperLTLtemp, the NBA $\A_{\K, \psi}^{\bowtie v}$  has a size that is bounded by a tower of exponentials of height depending linearly on the alternation depth of $\psi$. Thanks to this technical result, we obtain as expected: 

\begin{corollary}\label{cor:traces=lassos} For all formulas $\psi$ of
    \HyperLTLtemp, Kripke structures $\K$, and lasso assignments $\Pi$ such that $\Pi(\pi) \in \Lassos(\K)$ for all free variables $\pi$ in $\psi$,
    $\hsem\psi\Pi {\Traces(\K)} = \hsem \psi \Pi {\Lassos(\K)}$.
\end{corollary}
\begin{proof}
    The proof goes by induction on the quantifier depth of the formula $\psi$. 
    
    If $\psi$ is non quantified, the result is trivial since no quantifications
    remains to be evaluated.

    If $\psi = \exists \pi \; \psi'$, by definition of the semantics of the existential quantifier, and then by induction hypothesis,
     we have $\hsem \psi\Pi {\Traces(\K)} \geq \hsem \psi \Pi {\Lassos(\K)}$.
     Suppose that the inequality is strict, and let $v\in (\hsem \psi \Pi {\Lassos(\K)}, \hsem \psi\Pi {\Traces(\K)})$. Since $\hsem \psi\Pi {\Traces(\K)}>v$, by Lemma~\ref{lem:orna++}.\ref{item:logic->auto},
     $t_\Pi$ is then accepted by $\A_{\K, \psi}^{> v}$. Now, by 
     Lemma~\ref{lem:orna++}.\ref{item:auto->logic}, as $\Pi$ is a lasso assignment,
     $\hsem {\psi} {\Pi}{\Lassos(\K)}\geq v$, which contradicts the hypothesis that $v > \hsem {\psi} {\Pi}{\Lassos(\K)}$.

    If $\psi = \forall \pi \; \psi'$, by definition and by induction hypothesis,
    we have $\hsem \psi\Pi {\Traces(\K)} \leq \hsem \psi \Pi {\Lassos(\K)}$.
    Suppose that the inequality is strict, and let $v\in (\hsem \psi \Pi {\Traces(\K)}, \hsem \psi\Pi {\Lassos(\K)})$. By Lemma~\ref{lem:orna++}.\ref{item:logic->auto},
    $t_\Pi$ is then accepted by $\A_{\K, \psi}^{< v}$. Now, by 
    Lemma~\ref{lem:orna++}.\ref{item:auto->logic}, as $\Pi$ is a lasso assignment, 
    $\hsem {\psi} {\Pi}{\Lassos(\K)}\leq v < \hsem {\psi} {\Pi}{\Lassos(\K)}$. Contradiction.
\end{proof}

In Example~\ref{ex:no-lasso}, notice that although all lassos are associated by $\varphi$ with a positive semantics, we indeed get $\chsem {\forall \pi \, \varphi}{\Lassos(\K)} = 0$ by a convergence phenomenon: for instance, for the discounted sequence $\eta_i = (1/2)^i$ and the family of lassos $\{ t_n = (p (\neg p)^n)^\omega \}_{n \in \mathbb{N}}$, it holds that 
 $\sem\varphi{t_n} = (1/2) ^n$ which obviously converges to $0$ when $n$ goes to $\infty$. In contrast, we have also $\chsem {\forall \pi \; \varphi}{\Traces(\K)} = 0$ as there exists (non-lasso) traces associated by $\varphi$ with a zero semantics, such as the trace $t_\infty$ 
from Example~\ref{ex:no-lasso}.

\subparagraph*{Issues relative to automata-based model checking}
In \cite{Almagor14}, the model checking problem amounts to deciding whether all traces of a Kripke structure $\K$ have a semantics with respect to an \LTLtemp formula $\varphi$ greater than or equal to a threshold $v$. This translates to our setting in deciding whether $\chsem{\forall \pi\; \varphi}{\TracesOf{\K}} \geq v$. There, the authors check this by testing emptiness of the NBA $\A^{<v}_{\K, \varphi}$: this is indeed correct because of item~2 of Proposition~\ref{prop:ltldisc_nba_precice}, since if $\A^{<v}_{\K, \varphi}$ would accept a lasso $t$, it would be such that $\sem{\varphi}t < v$.

In the setting of \HyperLTLtemp, this leads to two difficulties. First, the second item of our Lemma~\ref{lem:orna++} is weaker than the one of Proposition~\ref{prop:ltldisc_nba_precice} as the comparison with the value $v$ is no longer strict, preventing the same argument as before in \LTLtemp. Second, our formulas contain $\forall$ and $\exists$ quantifiers (whose semantics are $\sup$ and $\inf$ operators respectively) and thus, we should be able to solve the model checking problems $\chsem{\psi}{\TracesOf{\K}} \geq v$ and $\chsem{\psi}{\TracesOf{\K}} \leq v$, for a closed formula $\psi$. Once again, the reasoning above fails in this case since it is not true a priori that $\chsem{\forall \pi\; \varphi}{\TracesOf{\K}} \leq v$ if and only if the NBA $\A^{>v}_{\K, \varphi}$ has an empty language. We demonstrate this more carefully in the example below. 

\begin{example}\label{ex:no-lasso-bis} We continue Example~\ref{ex:no-lasso}.
    Consider $\varphi_\pi = \G(\F_\eta p_\pi)\lor \F\G \lnot p_\pi$, and $\psi
    = \forall \pi \; \varphi_\pi$. As we discussed, for a trivial Kripke
    structure $\K$ allowing every trace, we have $\chsem\psi {\Traces(\K)} =
    \chsem \psi {\Lassos(\K)}=0$: there exist witnesses $t\in \Traces(\K)$
    where $\varphi_\pi$ evaluates to~$0$, but no such witness is a lasso of
    $\K$. For this formula, assume we would like to solve the model checking problem
    $\chsem\psi {\Traces(\K)}\leq 0$ (which indeed implies that $\chsem\psi
    {\Traces(\K)}=0$). Following an automata-based approach for this problem and  
    thus appealing to Lemma~\ref{lem:orna++}, this would require to build 
    the NBA $\A_{\K, \psi}^{>0}$ to check that its language is indeed empty. This automaton could be obtained by computing first the NBA
    $\A_{\K, \phi_\pi}^{>0}$ that accepts all lassos, complementing it
    (obtaining then an NBA with empty language) before taking its intersection
    with the traces generated by~$\K$, then projecting away the trace $\pi$,
    and finally, complementing it again. The so-obtained NBA would have unfortunately a non-empty language 
    and would accept the trace $t_{[\,]}$: such a model checking algorithm would thus wrongly conclude 
    that $\chsem\psi {\Traces(\K)} >0$.
  \qed
\end{example}

We therefore cannot directly rely on the construction of an NBA to fully and exactly model check \HyperLTLtemp. However, we propose two kinds of solutions in the rest of this section based on automata. First, we relax the problem, solving a \emph{promise} version of the model checking, following the terminology introduced in \cite{EVEN1984159}; second, we study several fragments for which the (exact) model checking task can be solved. 

\subparagraph*{Promise model checking}

We weaken the model checking problem to only solve a promise problem instead, where we allow our algorithm to answer arbitrarily in case of equality\footnote{With respect to promise problems of \cite{EVEN1984159}, the promise is that the value of the formula on the Kripke structure is different from the threshold.}: 


\begin{definition}
Given a Kripke structure $\K$, a closed \HyperLTLtemp formula $\psi$, a rational threshold $v\in[0,1]$, the \emph{model checking of} \HyperLTLtemp \emph{with non-equality promise} (associated with the exact problem $\chsem\psi{\Traces(\K)}\geq v$) answers positively when $\chsem\psi{\Traces(\K)}> v$, negatively when $\chsem\psi{\Traces(\K)}< v$, and arbitrarily when $\chsem\psi{\Traces(\K)}=v$.
\end{definition}

We can consider alternatively the symmetric question associated with the exact question $\chsem\psi{\Traces(\K)}\leq v$.

\begin{theorem}
\label{thm:MCHyperLTLDiscpromise}
The model checking of \HyperLTLtemp with non-equality promise is decidable with a non-elementary complexity.
\end{theorem}
\begin{proof}
   Let $\K$ be a Kripke structure, $\psi$ be  a closed \HyperLTLtemp formula, and $v\in[0,1]$ a rational threshold. 
   Thanks to Corollary~\ref{cor:traces=lassos}, note that $\chsem\psi{\Traces(\K)} = \chsem\psi{\Lassos(\K)}$. Consider then the NBA $\A_{\K, \psi}^{< v}$ of Lemma~\ref{lem:orna++}. Note that, since $\psi$ is closed, it runs on trace assignments over no traces: it either accepts the single infinite trace $t_{[]}$ that is a lasso (and thus, it is not empty), or it rejects it (and then, it is empty). 
The model checking algorithm with non-equality promise amounts to test emptiness for $\A_{\K, \psi}^{< v}$. 
Indeed, if $\chsem\psi{\Traces(\K)}> v$, i.e.~$\chsem\psi{\Lassos(\K)}> v$, we have $\chsem\psi{\Lassos(\K)}\leq v$ for no lassos, and thus (by property~2 of Lemma~\ref{lem:orna++}) the NBA $\A_{\K, \psi}^{< v}$ cannot accept any lasso: the emptiness check of this NBA declares that its language is empty and the algorithm answers positively.  
Symmetrically, if $\chsem\psi{\Traces(\K)}< v$, i.e.~$\chsem\psi{\Lassos(\K)}< v$, we have $\chsem\psi{\Lassos(\K)}\leq v$, and thus the single infinite lasso fulfils that; then, by item~2 of Lemma~\ref{lem:orna++}, it is accepted by the NBA $\A_{\K, \psi}^{< v}$: the emptiness check of this NBA declares that its language is non-empty. Hence, the algorithm answers negatively.

The non-elementary complexity follows from that fact that the NBA $\A_{\K, \psi}^{< v}$ has a size that is bounded by a tower of exponentials of height depending linearly on the alternation depth of $\psi$.
\end{proof}
 
We now show in the next two subsections the fragments of $\HyperLTLtemp$ for which one can answer the (exact) model checking problem.

\subparagraph*{Model checking of the positive and negative fragments}

We propose two first fragments for which we show the decidability of model checking. One is the fragment where negation is no longer allowed in \HyperLTLtemp formulas above $\U_\eta$ operator. For this reason, we call this restriction the \emph{positive fragment}. We consider as well its dual that we call the \emph{negative fragment}.

\begin{definition}
 The positive fragment of \HyperLTLtemp, denoted \HyperLTLtemppos,  is the subset of formulas $\psi$ described by the grammar (where $p\in \AP$, $\pi \in \mathcal V$, and $\eta\in \discfuns$):
 \begin{align*}
     \psi& ::= \exists\pi\, \psi \mid \forall\pi\, \psi \mid \phi \\
    \phi& ::= \beta \mid \phi \vee \phi \mid \phi \land \phi \mid \X\phi \mid \phi\U\phi \mid \phi\U_\eta\phi\\
     \beta& ::= p_\pi \mid \neg\beta \mid \beta \vee \beta \mid \X\beta \mid \beta\U\beta \tag*{(Boolean \LTL formulas)}
 \end{align*}
 
Symmetrically, we can define the fragment \HyperLTLtempneg obtained by considering negations of formulas of \HyperLTLtemppos. By pushing negations, this corresponds to the syntax 
\begin{align*}
    \psi& ::= \exists\pi\, \psi \mid \forall\pi\, \psi \mid \phi  \\
   \phi &::= \beta \mid \phi \vee \phi \mid \phi \land \phi \mid \X\phi \mid \phi\R\phi \mid \phi\R_\eta\phi\\
    \beta& ::= p_\pi \mid \neg\beta \mid \beta \vee \beta \mid \X\beta \mid \beta\U\beta \tag*{(Boolean \LTL formulas)}
\end{align*}
where the semantics of the \emph{release} operators $\R$ and $\R_\eta$ as obtained by a negation of the until operators, that is 
$$ \hsem{\phi_1 \R \phi_2} \Pi T =  \hsem{\neg(\neg \phi_1 \U \neg \phi_2)} \Pi T \qquad 
\hsem{\phi_1 \R_{\eta} \phi_2} \Pi T  = \hsem{\neg(\neg \phi_1 \U_{\eta} \neg \phi_2)} \Pi T
$$
\end{definition}

The two fragments are obviously related is the sense that for all closed formulas $\psi^+$ from \HyperLTLtemppos (respectively,~$\psi^-$ from \HyperLTLtempneg), there exists a formula 
$\psi^-$ from \HyperLTLtempneg (respectively,~$\psi^+$ from \HyperLTLtemppos) such that $\hsem{\psi^+}\Pi T = 1- \hsem{\psi^-}\Pi T$. 

The two model checking problems we address for \HyperLTLtemppos (respectively,~\HyperLTLtempneg) are the following ones: 
\begin{itemize}
    \item given a closed formula $\psi$ of \HyperLTLtemppos (respectively,~\HyperLTLtempneg), a Kripke structure $\K$ and a threshold $v\in [0, 1]$, decide if $\chsem\psi{\TracesOf{\K}} \geq v$
    (respectively, $\chsem\psi{\TracesOf{\K}} \leq v$);
    \item given a closed formula $\psi$ of \HyperLTLtemppos (respectively,~\HyperLTLtempneg), a Kripke structure $\K$ and a threshold\footnote{Notice that the threshold $0$ is not allowed here, to avoid convergence issues.} $v\in (0, 1]$  (respectively,~$v\in [0, 1)$), decide if $\chsem\psi{\TracesOf{\K}} \leq v$ (respectively,~$\chsem\psi{\TracesOf{\K}} \geq v$). 
\end{itemize}


The formula $\varphi_{\mathsf{OD}}^{\mathrm{temp}}$ of Example~\ref{ex:OD-disc} is in the fragment \HyperLTLtempneg. Indeed, the negation of the formula
can be rewritten as 
$$\psi := \exists \pi\, \exists \pi'\ \  \lowequiv \pi {\pi'} \land \F_\eta \lnot \lowequiv \pi {\pi'}$$ 
where $\lowequiv \pi {\pi'}$ is a Boolean \LTL formula, and the operator $\F_\eta$ does not involve any negation. This latter formula belongs to \HyperLTLtemppos. To model check the formula $\varphi_{\mathsf{OD}}^{\mathrm{temp}}$ against a threshold $\geq v$ (respectively,~$\leq v$), it suffices to model check the formula $\psi$ of \HyperLTLtemppos against the threshold $\leq 1-v$ (respectively,~$\geq 1-v$). 
Similarly, the formula $\varphi_{\mathsf{NI}}^{\mathrm{temp}}$ of Example~\ref{ex:noninference-disc} is in the fragment \HyperLTLtempneg.

The model checking algorithms for these fragments will rely on the model checking algorithm with non-equality promise together with a major ingredient, that is the characterisation of the set of possible satisfaction values generated by a positive formula, helping us control the convergence phenomena in infima/suprema of the semantics. 

\begin{definition}
\label{def:valpositive}
For all $k\in \N$, finite sets of discounted sequences $\Discounts \subseteq \discfuns$, and finite sets of weights $\weights\subset[0,1]$, the infinite set of values 
$\Valpositive{k}{\Discounts}{\weights}$ is given by 
%
$$\{0, 1\} \cup \weights\cup \{w\times \prod_{\ell=1}^{k'}\eta^{(\ell)}_{i_\ell}\mid w\in \weights, k'\leq k, 
(i_1, \ldots , i_{k'})  \in \mathbf N^{k'}, (\eta^{(1)}, \ldots, \eta^{(k')}) \in \Discounts^{k'}\}$$
\end{definition}

This set has the following properties, using as natural hypotheses that we only consider computable discounted sequences $\eta$:
\begin{restatable}{lemma}{lemtechnical}
    \label{lem:technical}
    For all $k\in \N$, finite sets $\Discounts \subseteq \discfuns$, and finite sets $\weights\subset[0,1]$,
    \begin{enumerate}
    \item for all $a\in (0, 1)$, the set $\Valpositive{k}{\Discounts}{\weights} \cap [a, 1]$ is finite and computable;
    \item every non-decreasing sequence of $\Valpositive{k}{\Discounts}{\weights}$ is stationary;
    \item every non-increasing sequence of $\Valpositive{k}{\Discounts}{\weights}$ either converges to 0 or is stationary.
    \end{enumerate}
\end{restatable}
\begin{proof}
    \begin{itemize}
        \item Let $a\in (0, 1)$. Since the functions $\eta\in H$ are decreasing and $H$ is finite, there is $i_0\in \mathbf N$ such that for all $i\geq i_0$ and $\eta\in \Discounts$, $\eta(i) < a$. Thus, $\Valpositive{k}{\Discounts}{\weights}\cap [a, 1] \subseteq \{1\} \cup \weights \allowbreak\cup \{w\times \prod_{\ell\leq k'}\eta_{i_\ell}\mid w\in \weights, k' \le k, \, \forall \ell\leq k \quad i_\ell \leq i_0, \eta \in \Discounts\}$ is finite and enumerable.
\item Let $(u_n)$ be a non-decreasing sequence of $\Valpositive{k}{\Discounts}{\weights}$. Thus, it has values in $\Valpositive{k}{\Discounts}{\weights}\cap [u_0, 1]$ that is finite, and thus is stationary.
\item Let $(u_n)$ be a non-increasing sequence of $\Valpositive{k}{\Discounts}{\weights}$. If the sequence does not converge to $0$, it has a positive lower bound $a$, and thus has values in $[a, 1]$. Since this set is finite, the sequence is stationary. \qedhere
    \end{itemize}
\end{proof}

Moreover, this set can indeed be used to obtain an over-approximation of the set of satisfaction values of a formula in \HyperLTLtemppos.

\begin{restatable}{lemma}{propHyperLTLtemppos}
    \label{lemma:prop-HyperLTLtemppos}
    Let $\psi$ be a formula of \HyperLTLtemppos, $k_\psi$ be the depth of $\U_\eta$ operators, $\Discounts_\psi$ be the finite set of functions $\eta$ appearing in these operators, and $\weights\subset[0,1]$ be a finite set. Then, for all $T\subseteq \Traces_\weights$, and $\Pi$ assignments to traces of $\Traces_\weights$, $\hsem\psi\Pi T \in 
    \Valpositive{k_\psi}{\Discounts_\psi}{\weights}$. 
\end{restatable}
\begin{proof}

    Notice that if $k\leq k'$ and $\Discounts \subseteq  \Discounts'$, then $\Valpositive{k}{\Discounts}{\weights} \subseteq \Valpositive{k'}{\Discounts'}{\weights}$. In particular, if $\psi$ 
    is a subformula of $\psi'$, we have $\Valpositive{k_\psi}{\Discounts_\psi}{\weights} \subseteq \Valpositive{k'}{\Discounts'}{\weights}$.
        Let $T\subseteq \Traces$. We show by induction that for all $\psi\in \HyperLTLtemppos$ and for all assignments $\Pi$, $\hsem\psi\Pi T \in \Valpositive{k_\psi}{\Discounts_\psi}{\weights}$. 
        \begin{itemize}
            \item For Boolean formulas $\beta$, $k_\beta = 0$ and $\Discounts_\beta = \emptyset$, and their semantics is in $\{0, 1\} = \Valpositive{k_\beta}{\Discounts_\beta}{\weights}$. 
            \item For unquantified formulas $\phi$ and $\phi'$, by induction, for all assignments $\Pi'$, we have $\hsem\phi{\Pi'} T \in \Valpositive{k_\phi}{\Discounts_\phi}{\weights}$ and 
            $\hsem{\phi'}{\Pi'} T \in \Valpositive{k_{\phi'}}{\Discounts_{\phi'}}{\weights}$. Then,  
             \begin{itemize}
              \item  $\hsem{\phi\lor \phi'}\Pi T  = \max(\hsem\phi\Pi T, \hsem{\phi'}\Pi T ) \in \Valpositive{k_{\phi}}{\Discounts_{\phi}}{\weights} \cup \Valpositive{k_{\phi'}}{\Discounts_{\phi'}}{\weights}
               \subseteq \Valpositive{k_{\phi\lor\phi'}}{\Discounts_{\phi\lor \phi'}}{\weights} $
                \item $\hsem{\phi\land \phi'}\Pi T  = \min(\hsem\phi\Pi T, \hsem{\phi'}\Pi T ) \in \Valpositive{k_{\phi}}{\Discounts_{\phi}}{\weights} \cup \Valpositive{k_{\phi'}}{\Discounts_{\phi'}}{\weights} 
                \subseteq \Valpositive{k_{\phi\land\phi'}}{\Discounts_{\phi\land \phi'}}{\weights}$
                \item $\hsem{\X\phi}\Pi T = \hsem\phi{\Pi[1, \infty]} T \in \Valpositive{k_{\phi}}{\Discounts_{\phi}}{\weights}  \subseteq \Valpositive{k_{\X\phi}}{\Discounts_{\X\phi}}{\weights}$
            \item 
            For until formulas, using again that $\Valpositive{k_{\phi}}{\Discounts_{\phi}}{\weights} \cup \Valpositive{k_{\phi'}}{\Discounts_{\phi'}}{\weights}   \subseteq 
              \Valpositive{k_{\phi\U\phi'}}{\Discounts_{\phi\U\phi'}}{\weights}$, 
            \[\hsem{\phi \U \phi'}\Pi T  = \sup_{i\geq 0}\big(\min(\underbrace{\hsem{\phi'} {\Pi[i,\infty]} T}_{\in \Valpositive{k_{\phi\U\phi'}}{\Discounts_{\phi\U\phi'}}{\weights}},\min_{0\leq j < i}\underbrace{\hsem{\phi}{\Pi[j,\infty]} T}_{\in \Valpositive{k_{\phi\U\phi'}}{\Discounts_{\phi\U\phi'}}{\weights}})\big) \]
            From item~3 of Lemma~\ref{lem:technical}, the minimum over $j$ is either 0 or is equal to one of the values, and thus belongs to $\Valpositive{k_{\phi\U\phi'}}{\Discounts_{\phi\U\phi'}}{\weights}$. Then, the (binary) minimum also belongs to $\Valpositive{k_{\phi\U\phi'}}{\Discounts_{\phi\U\phi'}}{\weights}$. From item~2 of Lemma~\ref{lem:technical}, the supremum is equal to one of the values, and thus belongs to $\Valpositive{k_{\phi\U\phi'}}{\Discounts_{\phi\U\phi'}}{\weights}$.
           \item 
            For discounted until formulas, by definition, 
            \[\hsem{\phi \U_\eta \phi'}\Pi T  = \sup_{i\geq 0}\big(\min(\eta_i\underbrace{\hsem{\phi'} {\Pi[i,\infty]} T}_{\in \Valpositive{k_{\phi'}}{\Discounts_{\phi'}}{\weights}},\min_{0\leq j < i}\eta_j\underbrace{\hsem{\phi}{\Pi[j,\infty]} T}_{\in \Valpositive{k_{\phi}}{\Discounts_{\phi}}{\weights}})\big) \]
            By definition, we have $k_{\phi\U_\eta\phi'} = \max(k_{\phi}, k_{\phi'})+1$ and $\Discounts_{\phi\U_\eta\phi'} = \Discounts_{\phi}\cup \Discounts_{\phi'}\cup\{\eta\}$. In particular, $\eta_i\hsem{\phi'} {\Pi[i,\infty]} T$ and $\eta_j\hsem{\phi}{\Pi[j,\infty]} T$ belongs to $\Valpositive{k_{\phi\U\phi'}}{\Discounts_{\phi\U\phi'}}{\weights}$. As before, from item~3 of Lemma~\ref{lem:technical}, the minimum over $j$ is either 0 or is equal to one of the values, and thus belongs to $\Valpositive{k_{\phi\U\phi'}}{\Discounts_{\phi\U\phi'}}{\weights}$. Then, the (binary) minimum also belongs to $\Valpositive{k_{\phi\U\phi'}}{\Discounts_{\phi\U\phi'}}{\weights}$. From item~2 of Lemma~\ref{lem:technical}, the supremum is equal to one of the values, and thus belongs to $\Valpositive{k_{\phi\U\phi'}}{\Discounts_{\phi\U\phi'}}{\weights}$.
         \end{itemize}
         
            \item Consider a formula $\exists \pi\, \psi$.
    By induction, for every assignment $\Pi$, and every trace $t\in T$,
    $\hsem\psi{\Pi[\pi\mapsto t]} T \in
    \Valpositive{k_\psi}{\Discounts_\psi}{\weights}$. By definition, $\hsem{\exists\pi\, \psi}{\Pi} T = \sup_{t\in T} \hsem\psi{\Pi[\pi\mapsto t]} T$. Consider a sequence $(t_n)_{n\in \N}$ of elements of $T$ such that $(\hsem\psi{\Pi[\pi\mapsto t_n]} T)_{n\in \N}$ is a non-decreasing sequence of $\Valpositive{k_\psi}{\Discounts_\psi}{\weights}$ converging towards $\hsem{\exists\pi\, \psi}{\Pi} T$. By item 2 of Lemma~\ref{lem:technical}, this sequence is stationary, and thus its limit $\hsem{\exists\pi\, \psi}{\Pi} T$ also belongs to $\Valpositive{k_\psi}{\Discounts_\psi}{\weights}$. A similar proof exists for formulas $\forall \pi \,\psi$, by using item 3 of Lemma~\ref{lem:technical}.
    \qedhere
        \end{itemize}
\end{proof}

This is enough to obtain an automata-based algorithm to model check \HyperLTLtemppos and \HyperLTLtempneg: 

\begin{theorem}\label{thm:HyperLTLtemp-MC}
    The model checking problems for \HyperLTLtemppos and \HyperLTLtempneg are decidable with a non-elementary complexity.
\end{theorem}
\begin{proof}
    Let $\psi$ be a closed formula of \HyperLTLtemppos, $\K$ be a Kripke structure. Let $\weights$ be the set of weights appearing in $\K$.
    \begin{itemize}
        \item Let $v\in [0, 1]$. To decide whether $\chsem \psi  {\Traces(\K)} \geq v$ (with
        $v\neq 0$ otherwise the answer is trivially true), we consider, by
        Lemma~\ref{lemma:prop-HyperLTLtemppos}, the greatest value $v'$ of $V_{k_\psi, H_\psi, \weights}$ smaller than $v$. This value can be computed as a search in the
        finite and enumerable set of possibles values $S = V_{k_\psi, H_\psi, \weights}\cap [a, 1]$ with $a = \sup_{i\mid \eta_i < v} \eta_i$ for any $\eta\in H_\psi$ (since the set $S$ contains $\eta_i$ and thus values smaller than $v$).
      %
%
We then let $v'' = (v'+v)/2$ and appeal to Theorem \ref{thm:MCHyperLTLDiscpromise}  and its algorithm with promise for the Kripke structure $\K$, the closed \HyperLTLtemp formula $\psi$ and the rational threshold $v''$; note that, as $v'' \notin V_{k_\psi, H_\psi, \weights}$, the promise is trivially satisfied. Now, on the one hand, if the algorithm with promise concludes that $\chsem \psi {\Traces(\K)} > v''$ then it holds that $\chsem \psi {\Traces(\K)} \ge v$. On the other hand, if the algorithm with promise concludes that $\chsem \psi {\Traces(\K)} < v''$, then it holds that $\chsem \psi {\Traces(\K)} < v$. 
        \item Let $v\in (0, 1]$. To decide whether $\chsem \psi {\Traces(\K)} \leq v$ (with
        $v\neq 1$, otherwise the answer is trivially true), we consider, by Lemma~\ref{lemma:prop-HyperLTLtemppos}, the lowest value $v'$ of
        $V_{k_\psi, H_\psi, \weights}$ greater than $v$: it exists (and can be computed) since $v$ is taken different from $0$. We then let $v'' = (v+v')/2$, and conclude the proof in this case symmetrically as in the previous case, by appealing to the algorithm from Theorem \ref{thm:MCHyperLTLDiscpromise} with the trivial promise that $\chsem \psi {\Traces(\K)} \neq v''$
        (as $v'' \notin V_{k_\psi, H_\psi, \weights}$). 
    \end{itemize}
    \noindent We obtain the decidability for \HyperLTLtempneg by some dual arguments.
\end{proof}

The complexity of this model checking algorithm is at least as high as
computing the automata given in Lemma~\ref{lem:orna++}, and thus at least
non-elementary (with respect to the quantifier depth of the formulas): this is
non-surprising since the model checking of \HyperLTL has already a
non-elementary lower bound in complexity. Formally, it also depends on the
search for the value $v'$ that is the greatest (respectively~lowest) value smaller
(respectively~greater) than a value~$v$ in a set of the form~$\Valpositive{k}{\Discounts}{\weights}$. This heavily
depends on the sequences $\eta$ in $\Discounts$. If we suppose that these sequences are
indeed geometrical, as it is the case for usual discounted automata where the
current weight is always multiplied by a power $\lambda^i$ of a discounted
parameter, this search is easy, and thus does not change the overall complexity of the
model checking algorithm.

\subparagraph*{Model checking of the alternation-free fragments}
The other two fragments that we consider, orthogonal to the previous ones, are the \emph{alternation-free} fragments: $\HyperLTLtempexists$ is the fragment containing formulas of the form $\exists \pi_1 \cdots \exists \pi_n \; \phi$ with $\phi$ quantifier-free, and $\HyperLTLtempforall$ is the fragment containing formulas of the form $\forall \pi_1 \cdots \forall \pi_n \; \phi$. 

The formula $\varphi_{\mathsf{OD}}^{\mathrm{temp}}$ describing observational determinism is in $\HyperLTLtempforall$ (as it contains simply two universal quantifications), but not the formula $\varphi_{\mathsf{NI}}^{\mathrm{temp}}$ describing non-inference (as it alternates between a universal quantification and an existential one). 

The model checking problems we consider for these two logics are asymmetric: 
given a closed formula $\psi$ of $\HyperLTLtempexists$ (respectively~$\HyperLTLtempforall$), a Kripke structure $\K$ and a threshold $v\in [0, 1]$, decide if $\chsem{\psi}{\Traces(\K)} \leq v$ (respectively~$\chsem{\psi}{\Traces(\K)} \geq v$).
These are tailored to mimic the model checking problem of \LTLtemp in \cite{Almagor14}, in order to correspond to the Boolean variants of classical model checking questions: for the existential fragment, the semantics of the existential quantification being a supremum over traces, asking whether the semantics of the formula is at most a certain threshold requires deciding if \emph{all} traces have an associated value at most the threshold; for the universal fragment, the semantics of the universal quantification being a infimum over traces, asking whether the semantics of the formula is at least a certain threshold requires deciding if \emph{all} traces have an associated value at least the threshold.
By a proof building upon Proposition~\ref{prop:ltldisc_nba_precice}, and similarly to what has been done in the proof of Theorem~\ref{thm:modelchecking_hyperLTL[F]}, we obtain: 

\begin{restatable}{theorem}{MCHyperLTLtempalternationfree}
    \label{thm:MC-HyperLTLtemp-alternationfree}
    The model checking of $\HyperLTLtempexists$ and $\HyperLTLtempforall$ is decidable with a non-elementary complexity.
\end{restatable}
\begin{proof}
    For the case of $\HyperLTLtempexists$, consider a closed formula $\psi = \exists\pi_1\cdots \exists \pi_n\; \phi$ with $\phi$ a quantifier-free formula, a Kripke structure $\K$ and a threshold $v\in [0,1]$. By Proposition~\ref{prop:ltldisc_nba_precice}, using the fact that $\phi$ can be seen as a \HyperLTLtemp formula, we build the NBA $\A_{\phi,>v}$. One then computes its intersection with $\A_{\K^n}$ to restrict ourselves to traces of $\Traces(\K^n)$, getting an NBA $\A_{\K, \phi}^{>v}$. Now, appealing to Lemma \ref{lem:traceproduct} and Remark \ref{rem:HyperLTLtemptoLTLtemp}, it holds that $\hsem\phi\Pi{\Traces(\K)} = \sem\phi{t_\Pi}$ for $t_\Pi \in \Lang(\A_{\K, \phi}^{>v})$. 
     This allows one to obtain that for every trace assignment $\Pi$, 
    \begin{enumerate}
        \item\label{item:1} if $\hsem\phi\Pi{\Traces(\K)}>v$ and every trace in $\Pi$ belongs to $\Traces(\K)$, then $\A_{\K, \phi}^{>v}$  accepts $t_\Pi$.
        \item\label{item:2} If $\A_{\K, \phi}^{>v}$ accepts $t_\Pi$ and $\Pi$ is a lasso assignment, then $\hsem\phi\Pi{\Traces(\K)}>v$ and every trace in $\Pi$ belongs to $\Traces(\K)$.
    \end{enumerate}
    We then project away all the components of $\A_{\K, \phi}^{>v}$ to mimic the existential quantifications, and let $\A$ be the so-obtained NBA over the alphabet containing only the letter $\emptyset$. We show that $L(\A) \neq \emptyset$ if and only if $\chsem{\psi}{\Traces(\K)}> v$. 

    Indeed, suppose that $L(\A)\neq \emptyset$. Due to projection, this implies that $L(\A_{\K, \phi}^{>v})\neq \emptyset$, and thus (by the properties of NBAs) the existence of a lasso assignment $\Pi$ where every trace belongs to $\Traces(\K)$ and $t_\Pi\in L(\A_{\K,\phi}^{ >v})$. By property~\ref{item:2} above, we deduce that $\hsem{\phi}{\Pi}{\Traces(\K)} >v$. The semantics of existential quantifiers being suprema, we obtain $\chsem{\psi}{\Traces(\K)} >v$. 

    Reciprocally, if $\chsem{\psi}{\Traces(\K)} >v$, there exist an assignment $\Pi$ (not a priori lassos) of traces in $\Traces(\K)$ such that $\hsem{\phi}{\Pi}{\Traces(\K)} > v$. By property~\ref{item:1} above, we obtain that $t_\Pi \in L(\A_{\K,\phi}^{>v})$. By projections, since every trace in $\Pi$ is in $\Traces(\K)$, this implies that $L(\A)\neq \emptyset$. 

    \medskip 
    For the case of $\HyperLTLtempforall$, notice that $\forall \pi_1 \cdots\forall \pi_n\; \phi$ is equivalent to $\lnot \psi$ with $\psi = \exists \pi_1\cdots \exists\pi_n \; \lnot \phi$. We thus apply the previous case to this formula with the threshold $1-v$. 
\end{proof}

\subparagraph*{Approximate model checking}

In \cite{Almagor14}, another solution is given for the model checking of \LTLtemp by means of some \LTLprop approximation: this is based on the approximation of the discounted sequences whose values are set to 0 when they become smaller than a given precision $\varepsilon$. This permits a translation from any \LTLtemp formula $\psi$ to some \LTLprop formula $\psi'$ with a close semantics at a  distance depending on $\psi$ and $\varepsilon$. This provides an approximate model checking algorithm. We recall formally this approach here and explain how to adapt it for $\HyperLTLtemp$ to obtain: 

\begin{theorem}\label{thm:approx-MC}
    For all $\Delta>0$ and formulas $\psi$ of \HyperLTLtemp, there exists a formula $\psi_{\Delta}$ of \HyperLTLprop such that for all $T$ and $\Pi$, $|\hsem{\psi}\Pi{T}-\hsem{\psi_{\Delta}}\Pi{T}|\leq \Delta$.
\end{theorem}

The formula $\psi_\Delta$ can be computed, and thus, from this theorem, we deduce an approximate model checking of \HyperLTLtemp: given a formula $\psi$ of \HyperLTLtemp, a rational threshold $v\in[0,1]$, a Kripke structure $\K$ and a precision $\Delta$, we answer ``yes'' if $\chsem{\psi}{\TracesOf{\K}}\geq v+\Delta$, ``no'' if $\chsem{\psi}{\TracesOf{\K}}\leq v-\Delta$, and anything otherwise. Because of the complexity of the model checking of \HyperLTLprop, this approximate algorithm has a non-elementary complexity.

\begin{proof}[Proof of Theorem~\ref{thm:approx-MC}]
    Let $n$ be the depth of operators $\U_\eta$ used in the formula $\psi$ of $\HyperLTLtemp$. We let $\varepsilon$ a rational number such that $\varepsilon^n \leq \Delta$. The idea is to replace each operator $\U_\eta$ in $\psi$ by a subformula of \HyperLTLprop that computes the original operator with an error bounded by $\varepsilon$. The accumulation of $n$ such errors will thus be bounded by $\Delta$.

    Consider thus a subformula of $\psi$ of the form $\psi_1 \U_\eta \psi_2$. First, we define a new "discounted" sequence $\eta^\varepsilon$ defined for all $i\in\N$ by $\eta^\varepsilon_i = \eta_i$ if $\eta_i>\varepsilon$, and $\eta^\varepsilon_i=0$ otherwise: we cut the sequence $\eta$ at some point, and make it flat to $0$ at the end. Note that $\eta^\varepsilon$ is not a legal discounted sequence since it is no longer (strictly) decreasing, but the semantics of formulas in \HyperLTLtemp can be defined even if $\eta$ is simply supposed to be non-increasing. 
    
    We replace in $\psi$ the subformula $\psi_1 \U_\eta \psi_2$ by $\psi_1 \U_{\eta^\epsilon} \psi_2$. The semantics of the subformula is modified by at most $\varepsilon$. It is thus easy to obtain by induction that applying this modification for every $\U_\eta$ subformula to get a formula  $\psi'$ leads to $|\hsem{\psi}\Pi{T}-\hsem{\psi'}\Pi{T}|\leq \varepsilon^n \leq  \Delta$, for all $T$ and $\Pi$.

    The last step is to see that each subformula $\psi_1 \U_{\eta^\epsilon} \psi_2$ with a stationary sequence $\eta^\varepsilon$ only considers a bounded prefix of length $\max\{n\mid \eta^\varepsilon_n >0\}$, and can thus be replaced by a formula of \HyperLTLprop where a discount factor is applied propositionally (by means of some operator of the form $\scalmult {\eta^\varepsilon_i}$) for each prefix up to the maximal length. This formula has a size linear in $\psi_1$ and $\psi_2$, and linear in $\max\{n\mid \eta^\varepsilon_n >0\}$.
\end{proof}




\section{Conclusion}

We have proposed a framework for qualitative reasoning about hyperproperties,
by considering formulas whose semantics range in $[0,1]$. We introduce two
logics extending $\HyperLTL$, namely $\HyperLTLprop$ and $\HyperLTLtemp$, and
study their model checking problem. For the former, we propose an algorithm
with a complexity similar to the one of $\HyperLTL$. For the latter, we propose
some model checking algorithms for fragments of this logic as well a promise
algorithm for the full logic, the promise being that the semantics of the formula 
on the Kripke structure is different from the input rational threshold. 
 We also extend an existing approximation algorithm for the full logic.  

We leave as open the decidability of the exact model checking problem for the
whole logic \HyperLTLtemp, which, thanks to the promise algorithm, boils down to decide whether 
the value of a formula over a Kripke structure is equal to a given threshold.

It is very tempting to combine the propositional quality reasoning of
\HyperLTLprop with the temporal quality reasoning of \HyperLTLtemp.
Unfortunately, even without hyperproperties, it is shown in
\cite[Theorem~6.6]{Almagor14} that the undecidability is obtained very quickly
for \LTLprop and \LTLtemp, already just in adding to \LTLtemp the operator
$\oplus_{1/2}$ that we introduce in Section~\ref{sec:prop-Defs}: as a
corollary, we can obtain easily that the combination of \HyperLTLprop and
\HyperLTLtemp leads to a logic with undecidable model checking, even already if
we simply add to \HyperLTLtemp the average operator $\oplus_{1/2}$.
However, some propositional operators are studied in \cite{Almagor14} that can
be added to \LTLtemp, while keeping some decidability: these operators can
model propositional qualities like competence, necessity, and confidence. We
leave as future works to study these extensions in the hyperproperty setting. Another question for future works concerns the promise version of the model checking problem for the combination of \HyperLTLprop and \HyperLTLtemp: the proof for \LTLprop and \LTLtemp in \cite[Theorem~6.6]{Almagor14} relies on an equality case that we cannot easily remove, and thus the question remains open.

As future works, we would also like to study qualitative extensions of
$\mathrm{HyperPCTL}$~\cite{10.1007/978-3-319-99154-2_2}, a temporal logic for
reasoning about probabilistic hyperproperties of discrete-time Markov chains. 

\bibliography{references}

\begin{thebibliography}{10}

\bibitem{10.1007/978-3-319-99154-2_2}
Erika {\'A}brah{\'a}m and Borzoo Bonakdarpour.
\newblock {HyperPCTL}: A temporal logic for probabilistic hyperproperties.
\newblock In {\em Quantitative Evaluation of Systems}. Springer, 2018.
\newblock \href {https://doi.org/10.1007/978-3-319-99154-2_2} {\path{doi:10.1007/978-3-319-99154-2_2}}.

\bibitem{Almagor14}
Shaull Almagor, Udi Boker, and Orna Kupferman.
\newblock Formally reasoning about quality.
\newblock {\em Journal of the ACM}, 63(3):1--56, 2016.
\newblock \href {https://doi.org/10.1145/2875421} {\path{doi:10.1145/2875421}}.

\bibitem{DBLP:journals/tocl/BolligGMZ14}
Benedikt Bollig, Paul Gastin, Benjamin Monmege, and Marc Zeitoun.
\newblock Pebble weighted automata and weighted logics.
\newblock {\em {ACM} Trans. Comput. Log.}, 15(2):15:1--15:35, 2014.
\newblock \href {https://doi.org/10.1145/2579819} {\path{doi:10.1145/2579819}}.

\bibitem{7588075}
Marco Chiesa, Guy Kindler, and Michael Schapira.
\newblock Traffic engineering with equal-cost-multipath: An algorithmic perspective.
\newblock {\em IEEE/ACM Transactions on Networking}, 25(2):779--792, 2017.
\newblock \href {https://doi.org/10.1109/TNET.2016.2614247} {\path{doi:10.1109/TNET.2016.2614247}}.

\bibitem{ClarkeEmerson81}
Edmund~M. Clarke and E.~Allen Emerson.
\newblock Design and synthesis of synchronization skeletons using branching-time temporal logic.
\newblock In {\em Proceedings of the Workshop on Logics of Programs}, volume 131 of {\em Lecture Notes in Computer Science}, pages 52--71. Springer, 1982.

\bibitem{HMC2020}
Edmund~M. Clarke, Thomas~A. Henzinger, Helmut Veith, and Roderick Bloem, editors.
\newblock {\em Handbook of Model Checking}.
\newblock Springer Cham, 2020.
\newblock \href {https://doi.org/10.1007/978-3-319-10575-8} {\path{doi:10.1007/978-3-319-10575-8}}.

\bibitem{ClarkFink14}
M.~R. Clarkson, Bernd Finkbeiner, M.~Koleini, K.~K. Micinski, Markus~N. Rabe, and C{\'e}sar S{\'a}nchez.
\newblock Temporal logics for hyperproperties.
\newblock In {\em {POST} 2014}, volume 8414 of {\em LNCS}, pages 265--284. Springer, 2014.
\newblock \href {https://doi.org/10.1007/978-3-642-54792-8_15} {\path{doi:10.1007/978-3-642-54792-8_15}}.

\bibitem{Hyperproperties}
M.~R. Clarkson and F.~Schneider.
\newblock Hyperproperties.
\newblock {\em Journal of Computer Security}, 18:1157--1210, 2010.
\newblock \href {https://doi.org/10.1109/CSF.2008.7} {\path{doi:10.1109/CSF.2008.7}}.

\bibitem{DBLP:journals/tcs/DrosteG07}
Manfred Droste and Paul Gastin.
\newblock Weighted automata and weighted logics.
\newblock {\em Theor. Comput. Sci.}, 380(1-2):69--86, 2007.
\newblock URL: \url{https://doi.org/10.1016/j.tcs.2007.02.055}, \href {https://doi.org/10.1016/J.TCS.2007.02.055} {\path{doi:10.1016/J.TCS.2007.02.055}}.

\bibitem{Spot}
Alexandre Duret-Lutz, Etienne Renault, Maximilien Colange, Florian Renkin, Alexandre Gbaguidi~Aisse, Philipp Schlehuber-Caissier, Thomas Medioni, Antoine Martin, J{\'e}r{\^o}me Dubois, Cl{\'e}ment Gillard, and Henrich Lauko.
\newblock From {S}pot 2.0 to {S}pot 2.10: What's new?
\newblock In {\em {CAV} 2022}, volume 13372 of {\em LNCS}, pages 174--187. Springer, 2022.
\newblock \href {https://doi.org/10.1007/978-3-031-13188-2_9} {\path{doi:10.1007/978-3-031-13188-2_9}}.

\bibitem{EVEN1984159}
Shimon Even, Alan~L. Selman, and Yacov Yacobi.
\newblock The complexity of promise problems with applications to public-key cryptography.
\newblock {\em Information and Control}, 61(2):159--173, 1984.
\newblock URL: \url{https://www.sciencedirect.com/science/article/pii/S001999588480056X}, \href {https://doi.org/10.1016/S0019-9958(84)80056-X} {\path{doi:10.1016/S0019-9958(84)80056-X}}.

\bibitem{FaeLeg08}
Marco Faella, Axel Legay, and Mari{\"e}lle Stoelinga.
\newblock Model checking quantitative linear time logic.
\newblock {\em Electronic Notes in Theoretical Computer Science}, 220:61--77, 2008.
\newblock \href {https://doi.org/10.1016/j.entcs.2008.11.019} {\path{doi:10.1016/j.entcs.2008.11.019}}.

\bibitem{FerLuo24}
Angelo Ferrando, Giulia Luongo, Vadim Malvone, and Aniello Murano.
\newblock Theory and practice of quantitative {ATL}.
\newblock In {\em PRIMA 2024: Principles and Practice of Multi-Agent Systems}, volume 15395 of {\em LNCS}. Springer, 2024.
\newblock \href {https://doi.org/10.1007/978-3-031-77367-9_18} {\path{doi:10.1007/978-3-031-77367-9_18}}.

\bibitem{DBLP:conf/vmcai/Finkbeiner21}
Bernd Finkbeiner.
\newblock Model checking algorithms for hyperproperties (invited paper).
\newblock In Fritz Henglein, Sharon Shoham, and Yakir Vizel, editors, {\em Verification, Model Checking, and Abstract Interpretation - 22nd International Conference, {VMCAI} 2021, Copenhagen, Denmark, January 17-19, 2021, Proceedings}, volume 12597 of {\em Lecture Notes in Computer Science}, pages 3--16. Springer, 2021.

\bibitem{DBLP:conf/concur/FinkbeinerH16}
Bernd Finkbeiner and Christopher Hahn.
\newblock Deciding hyperproperties.
\newblock In {\em {CONCUR} 2016}, volume~59 of {\em LIPIcs}, pages 13:1--13:14. Schloss Dagstuhl - Leibniz-Zentrum f{\"{u}}r Informatik, 2016.
\newblock \href {https://doi.org/10.4230/LIPICS.CONCUR.2016.13} {\path{doi:10.4230/LIPICS.CONCUR.2016.13}}.

\bibitem{DBLP:journals/corr/abs-1905-13514}
Bernd Finkbeiner, Christopher Hahn, and Hazem Torfah.
\newblock Model checking quantitative hyperproperties.
\newblock In {\em {CAV} 2018}, volume 10981 of {\em LNCS}. Springer, 2018.
\newblock \href {https://doi.org/10.1007/978-3-319-96145-3_8} {\path{doi:10.1007/978-3-319-96145-3_8}}.

\bibitem{DBLP:conf/stacs/Finkbeiner017}
Bernd Finkbeiner and Martin Zimmermann.
\newblock The first-order logic of hyperproperties.
\newblock In {\em 34th Symposium on Theoretical Aspects of Computer Science, {STACS} 2017, March 8-11, 2017, Hannover, Germany}, volume~66 of {\em LIPIcs}, pages 30:1--30:14. Schloss Dagstuhl - Leibniz-Zentrum f{\"{u}}r Informatik, 2017.

\bibitem{info-flow}
R.~Focardi and R.~Gorrieri.
\newblock Classification of security properties (part {I}: Information flow).
\newblock In {\em {FOSAD} 2000}, volume 2171 of {\em LNCS}, pages 331--396. Springer, 2000.
\newblock \href {https://doi.org/10.1007/3-540-45608-2_6} {\path{doi:10.1007/3-540-45608-2_6}}.

\bibitem{ltl-modelchecking}
Rob Gerth, Doron Peled, Moshe~Y. Vardi, and Pierre Wolper.
\newblock Simple on-the-fly automatic verification of linear temporal logic.
\newblock In {\em {PSTV} 1995}, pages 3--15. Springer, 1995.
\newblock \href {https://doi.org/10.1007/978-0-387-34892-6_1} {\path{doi:10.1007/978-0-387-34892-6_1}}.

\bibitem{GogMes82}
Joseph~A. Goguen and Jos{\'e} Meseguer.
\newblock Security policies and security models.
\newblock In {\em IEEE Symposium on Security and Privacy}, pages 11--20. {IEEE}, 1982.
\newblock \href {https://doi.org/10.1109/SP.1982.10014} {\path{doi:10.1109/SP.1982.10014}}.

\bibitem{GraMon26}
Samuel Graepler, Benjamin Monmege, and Jean-Marc Talbot.
\newblock {Reasoning About Quality in Hyperproperties}.
\newblock In Stefano Guerrini and Barbara K\"{o}nig, editors, {\em 34th EACSL Annual Conference on Computer Science Logic (CSL 2026)}, volume 363 of {\em Leibniz International Proceedings in Informatics (LIPIcs)}, pages 45:1--45:18, Dagstuhl, Germany, 2026. Schloss Dagstuhl -- Leibniz-Zentrum f{\"u}r Informatik.
\newblock \href {https://doi.org/10.4230/LIPIcs.CSL.2026.45} {\path{doi:10.4230/LIPIcs.CSL.2026.45}}.

\bibitem{Spin}
Gerard~J. Holzmann.
\newblock The model checker {SPIN}.
\newblock {\em {IEEE} Trans. Software Eng.}, 23(5):279--295, 1997.
\newblock \href {https://doi.org/10.1109/32.588521} {\path{doi:10.1109/32.588521}}.

\bibitem{DBLP:conf/sp/McLean94}
John McLean.
\newblock A general theory of composition for trace sets closed under selective interleaving functions.
\newblock In {\em Symposium on Research in Security and Privacy}, pages 79--93. {IEEE} Computer Society, 1994.
\newblock \href {https://doi.org/10.1109/RISP.1994.296590} {\path{doi:10.1109/RISP.1994.296590}}.

\bibitem{Shannon49}
C.~E. Shannon.
\newblock The synthesis of two terminal switching circuits.
\newblock {\em The Bell System Technical Journal}, 28(1):59--98, 1949.
\newblock \href {https://doi.org/10.1002/j.1538-7305.1949.tb03624.x} {\path{doi:10.1002/j.1538-7305.1949.tb03624.x}}.

\bibitem{SisCla85}
A.~P. Sistla and Edmund~M. Clarke.
\newblock The complexity of propositional linear temporal logic.
\newblock {\em Journal of the ACM}, 32:733--749, 1985.
\newblock \href {https://doi.org/10.1145/3828.3837} {\path{doi:10.1145/3828.3837}}.

\bibitem{minentropy}
G.~Smith.
\newblock On the foundations of quantitative information flow.
\newblock In {\em {FoSSaCS} 2009}, volume 5504 of {\em LNCS}. Springer, 2009.
\newblock \href {https://doi.org/10.1007/978-3-642-00596-1_21} {\path{doi:10.1007/978-3-642-00596-1_21}}.

\bibitem{10.1007/11547662_24}
Tachio Terauchi and Alex Aiken.
\newblock Secure information flow as a safety problem.
\newblock In Chris Hankin and Igor Siveroni, editors, {\em Static Analysis}, pages 352--367, Berlin, Heidelberg, 2005. Springer Berlin Heidelberg.
\newblock \href {https://doi.org/10.1007/11547662_24} {\path{doi:10.1007/11547662_24}}.

\bibitem{DBLP:journals/tcs/YasuokaT14}
Hirotoshi Yasuoka and Tachio Terauchi.
\newblock Quantitative information flow as safety and liveness hyperproperties.
\newblock {\em Theoretical Computer Science}, 538:167--182, 2014.
\newblock \href {https://doi.org/10.1016/J.TCS.2013.07.031} {\path{doi:10.1016/J.TCS.2013.07.031}}.

\bibitem{601322}
A.~Zakinthinos and E.S. Lee.
\newblock A general theory of security properties.
\newblock In {\em Proceedings. 1997 IEEE Symposium on Security and Privacy (Cat. No.97CB36097)}, pages 94--102, 1997.
\newblock \href {https://doi.org/10.1109/SECPRI.1997.601322} {\path{doi:10.1109/SECPRI.1997.601322}}.

\bibitem{DBLP:conf/csfw/ZdancewicM03}
Steve Zdancewic and Andrew~C. Myers.
\newblock Observational determinism for concurrent program security.
\newblock In {\em {CSFW} 2003}. {IEEE} Computer Society, 2003.
\newblock \href {https://doi.org/10.1109/CSFW.2003.1212703} {\path{doi:10.1109/CSFW.2003.1212703}}.

\end{thebibliography}

\end{document}